\documentclass[submission,copyright,creativecommons]{eptcs}
\usepackage[T1]{fontenc}
\usepackage[utf8]{inputenc}
\usepackage[USenglish]{babel}
\usepackage{isabelle}
\usepackage{xcolor}
\usepackage{isabellesym}
\usepackage{amsmath}
\usepackage{amssymb}
\usepackage[numbers, sort&compress, sectionbib]{natbib}
\IfFileExists{DOF-core.sty}{}{%
  \PackageError{DOF-core}{Isabelle/DOF not installed. 
    This is a Isabelle_DOF project. The document preparation requires
    the Isabelle_DOF framework. Please obtain the framework by cloning
    the Isabelle_DOF git repository, i.e.: 
         "git clone https://git.logicalhacking.com/Isabelle_DOF/Isabelle_DOF"
    You can install the framework as follows:
         "cd Isabelle_DOF/document-generator && ./install"}{%
    For further help, see https://git.logicalhacking.com/Isabelle_DOF/Isabelle_DOF}
}

\newcommand{\subtitle}[1]{%
  \PackageError{DOF-eptcs-UNSUPPORTED}
               {The LaTeX class eptcs does not support subtitles.}
               {}\stop%
}%

\RequirePackage{DOF-core}
\newcommand{\dofeptcsinstitute}[1]{\mbox{}\\\protect\scriptsize%
  \protect\begin{tabular}[t]{@{\protect\footnotesize}c@{}}%
    #1%
  \protect\end{tabular}%
}

\makeatletter
\newisadof{text.scholarly_paper.author}%
[label=,type=%
,scholarly_paper.author.email=%
,scholarly_paper.author.affiliation=%
,scholarly_paper.author.orcid=%
,scholarly_paper.author.http_site=%
][1]{%
  \protected@write\@auxout{}{%
  \string\protect\string\addauthor{%
   #1 %
    \string\mbox{}\string\dofeptcsinstitute{\commandkey{scholarly_paper.author.affiliation}} %
    \string\mbox{}\string\email{\commandkey{scholarly_paper.author.email}} %
  }
  }
}
\makeatother
\usepackage{DOF-core}
\usepackage{DOF-scholarly_paper}

\IfFileExists{preamble.tex}{
\usepackage{textcomp}
\usepackage{xcolor}
\usepackage{paralist}
\usepackage{datetime2}      
\usepackage{amssymb}
\usepackage{listings}
\usepackage{lstisadof}
\usepackage{xspace}
\usepackage[draft]{fixme}
\usepackage{wrapfig}

\providecommand{\event}{F-IDE 2019} 

\def\titlerunning{Deeply Integrating C11 Code Support into Isabelle/PIDE} 
\def\authorrunning{F. Tuong and B. Wolff} 

\renewcommand{\DOFinstitute}{}
\title{Deeply Integrating C11 Code Support\\ into Isabelle/PIDE} 

\newcommand{\house}{\mathrel{\substack{\wedge\\[-.2em]\sqcup}}}
}{}%
\begin{document}

\maketitle
\begin{isabellebody}%
\setisabellecontext{paper}%
\isadelimtheory
\endisadelimtheory
\isatagtheory
\endisatagtheory
{\isafoldtheory}%
\isadelimtheory
\endisadelimtheory
\isadelimML
\endisadelimML
\isatagML
\endisatagML
{\isafoldML}%
\isadelimML
\endisadelimML
\begin{isamarkuptext*}%
[label = {fred},type = {scholarly_paper.author}, args={label = {fred},type = {scholarly_paper.author}, scholarly_paper.author.email = {ftuong@lri.fr}, scholarly_paper.author.affiliation = {LRI, Univ. Paris-Sud, CNRS, Université Paris-Saclay}, scholarly_paper.author.http_site = {https://www.lri.fr/~ftuong/}, scholarly_paper.author.orcid = {}}]Frédéric Tuong%
\end{isamarkuptext*}\isamarkuptrue%
\begin{isamarkuptext*}%
[label = {bu},type = {scholarly_paper.author}, args={label = {bu},type = {scholarly_paper.author}, scholarly_paper.author.email = {wolff@lri.fr}, scholarly_paper.author.affiliation = {LRI, Univ. Paris-Sud, CNRS, Université Paris-Saclay}, scholarly_paper.author.http_site = {https://www.lri.fr/~wolff/}, scholarly_paper.author.orcid = {}}]Burkhart Wolff%
\end{isamarkuptext*}\isamarkuptrue%
\begin{isamarkuptext*}%
[label = {abs},type = {scholarly_paper.abstract}, args={label = {abs},type = {scholarly_paper.abstract}, scholarly_paper.abstract.keywordlist = {{User Interface, Integrated Development, Program Verification, Shallow Embedding}}}]We present a framework for C code in C11 syntax deeply integrated into the Isabelle/PIDE
  development environment. Our framework provides an abstract interface for verification back-ends
  to be plugged-in independently. Thus, various techniques such as deductive program verification or
  white-box testing can be applied to the same source, which is part of an integrated PIDE document
  model. Semantic back-ends are free to choose the supported C fragment and its semantics. In
  particular, they can differ on the chosen memory model or the specification mechanism
  for framing conditions.

  Our framework supports semantic annotations of C sources in the form of comments. Annotations
  serve to locally control back-end settings, and can express the term focus to which
  an annotation refers. Both the logical and the syntactic context are available when semantic
  annotations are evaluated. As a consequence, a formula in an annotation can refer both 
  to  HOL or C variables.

  Our approach demonstrates the degree of maturity and expressive power the Isabelle/PIDE subsystem
  has achieved in recent years. Our integration technique employs Lex and Yacc style grammars to 
  ensure efficient deterministic parsing. We present two case studies for the integration of 
  (known) semantic back-ends in order to validate the design decisions for our
  back-end interface.%
\end{isamarkuptext*}\isamarkuptrue%
\begin{isamarkupsection*}%
[label = {intro},type = {scholarly_paper.introduction}, args={label = {intro},type = {scholarly_paper.introduction}, Isa_COL.text_element.level = {}, Isa_COL.text_element.referentiable = {False}, Isa_COL.text_element.variants = {{STR ''outline'', STR ''document''}}, scholarly_paper.text_section.main_author = {}, scholarly_paper.text_section.fixme_list = {}, Isa_COL.text_element.level = {}}]Introduction%
\end{isamarkupsection*}\isamarkuptrue%
\begin{isamarkuptext*}%
[label = {introtext},type = {scholarly_paper.introduction}, args={label = {introtext},type = {scholarly_paper.introduction}, Isa_COL.text_element.level = {}, Isa_COL.text_element.referentiable = {False}, Isa_COL.text_element.variants = {{STR ''outline'', STR ''document''}}, scholarly_paper.text_section.main_author = {}, scholarly_paper.text_section.fixme_list = {}, Isa_COL.text_element.level = {}}]\noindent{}Recent successes like the Microsoft Hypervisor project \cite{DBLP:conf/fm/LeinenbachS09},
the verified CompCert compiler \cite{DBLP:journals/cacm/Leroy09}
and the seL4 microkernel \cite{DBLP:conf/sosp/KleinEHACDEEKNSTW09,DBLP:journals/tocs/KleinAEMSKH14} 
show that the verification of low-level systems code has become feasible.
However, a closer look at the underlying verification engines  
VCC \cite{DBLP:conf/tphol/CohenDHLMSST09}, 
or Isabelle/AutoCorres \cite{DBLP:conf/pldi/GreenawayLAK14}
show that the road is still bumpy: the  empirical cost evaluation  of the L4.verified project 
\cite{DBLP:journals/tocs/KleinAEMSKH14} reveals that a very substantial part  of the overall 
effort of about one third of the 28 man years went into the development of libraries and the 
associated tool-chain. Accordingly, the project authors \cite{DBLP:journals/tocs/KleinAEMSKH14} 
express the hope that these overall investments will not have to be repeated for 
``similar projects''.

In fact, none of these verifying compiler tool-chains capture all aspects of ``real life'' 
programming languages such as C. The variety of supported language fragments seem to contradict 
the assumption that we will all converge to one comprehensive tool-chain soon. There are so many 
different choices concerning memory models, non-standard control flow, and execution models 
that a generic framework is desirable: in which verified compilers, deductive verification, 
static analysis and test techniques (such as \cite{DBLP:conf/tap/Keller18}, 
\cite{DBLP:conf/itp/AissatVW16}) can be developed and used inside the Isabelle platform
as part of an integrated document.

In this paper we present Isabelle/C~\footnote{The current developer snapshot is provided in
  \url{https://gitlri.lri.fr/ftuong/isabelle_c}.}, a generic framework in
spirit similar to Frama-C \cite{frama-c-home-page}. In contrast to the latter, Isabelle/C is
deeply integrated into the Isabelle/PIDE document model \cite{DBLP:conf/itp/Wenzel14}. Based on
the C11 standard (ISO/IEC 9899:2011), Isabelle/C parses C11 code inside a rich IDE supporting static
scoping. SML user-programmed extensions can benefit from the parallel evaluation techniques of
Isabelle. The plug-in mechanism of Isabelle/C can integrate diverse semantic representations,
including those already made available in Isabelle/HOL \cite{DBLP:books/sp/NipkowPW02}: AutoCorres
\cite{DBLP:conf/pldi/GreenawayLAK14}, IMP2 \cite{DBLP:journals/afp/LammichW19},
Orca \cite{bockenek:hal-02069705}, or Clean (discussed in this paper). A particular advantage of
the overall approach compared to systems like Frama-C or VCC is that all these semantic
theories are conservative extensions of HOL, hence no axiom-generators are used that produce the
"background theory" and the verification conditions passed to automated provers. Isabelle/C provides
a general infrastructure for semantic annotations specific for back-ends, i.e. modules that generate
from the C source a set of definitions and derive automatically theorems over them.
Last but not least, navigation features of annotations make the logical context explicit in which 
theorems and proofs are interpreted.%
\end{isamarkuptext*}\isamarkuptrue%
\begin{isamarkupfigure*}%
[label = {C-sample},type = {Isa_COL.figure}, args={label = {C-sample},type = {Isa_COL.figure}, Isa_COL.figure.relative_width = {60}, Isa_COL.figure.src = {figures/A-C-Source}, Isa_COL.figure.spawn_columns = {True}}]A C11 sample in Isabelle/jEdit%
\end{isamarkupfigure*}\isamarkuptrue%
\begin{isamarkuptext}%
The heart of Isabelle/C, the new \isa{\isacommand{C}{\isacartoucheopen}\ {\isachardot}{\isachardot}\ {\isacartoucheclose}} command, is shown in \csname isaDof.ref\endcsname[type={Isa_COL.figure}]{C-sample}. 
Analogously to the existing \isa{\isacommand{ML}{\isacartoucheopen}\ {\isachardot}{\isachardot}\ {\isacartoucheclose}} command, it allows for editing C
sources inside the \isa{{\isacartoucheopen}\ {\isachardot}{\isachardot}\ {\isacartoucheclose}} brackets, where C code is
parsed on the fly in a ``continuous check, continuous build'' manner. A parsed source is coloured
according to the usual conventions applying for Isabelle/HOL variables and keywords. A static
scoping analysis makes the bindings inside the source explicit such that editing gestures like
hovering and clicking may allow the user to reveal the defining variable occurrences and C type
information (see yellow sub-box in the screenshot \csname isaDof.ref\endcsname[type={Isa_COL.figure}]{C-sample}). The C source
may contain comments to set up semantic back-ends. Isabelle/C turns out to be sufficiently efficient
for C sources such as the seL4 project.%
\end{isamarkuptext}\isamarkuptrue%
\begin{isamarkuptext}%
This paper proceeds as follows: in \csname isaDof.ref\endcsname[type={text}]{background}, we
briefly introduce Isabelle/PIDE and its document model, into which our framework is integrated. In
\csname isaDof.ref\endcsname[type={text}]{frontend_arch} and
\csname isaDof.ref\endcsname[type={text}]{ctests}, we discuss the build process and present some experimental results 
on the integrated parser. The handling of semantic annotations comments --- a vital part for 
back-end developers --- is discussed in \csname isaDof.ref\endcsname[type={text}]{annotations}, while in
\csname isaDof.ref\endcsname[type={text}]{backends} we present some techniques to integrate back-ends
into our framework at the hand of examples.%
\end{isamarkuptext}\isamarkuptrue%
\begin{isamarkupsection*}%
[label = {background},type = {scholarly_paper.technical}, args={label = {background},type = {scholarly_paper.technical}, Isa_COL.text_element.level = {}, Isa_COL.text_element.referentiable = {False}, Isa_COL.text_element.variants = {{STR ''outline'', STR ''document''}}, scholarly_paper.text_section.main_author = {}, scholarly_paper.text_section.fixme_list = {}, Isa_COL.text_element.level = {}, scholarly_paper.technical.definition_list = {}}]Background: PIDE and the Isabelle Document Model%
\end{isamarkupsection*}\isamarkuptrue%
\begin{isamarkuptext}%
\noindent{}The Isabelle system is based on a generic document model allowing for
efficient, highly-parallelized evaluation and checking of its document content (cf. 
\cite{DBLP:conf/itp/Wenzel14,DBLP:journals/corr/Wenzel14,DBLP:conf/mkm/BarrasGHRTWW13}
for the fairly innovative technologies underlying the Isabelle architecture).
These technologies allow for scaling up to fairly large  documents: we have seen documents
with 150 files be loaded in about 4 min, and individual files
--- like the x86 model generated from Antony Fox' L3 specs --- have 80~kLoC and were loaded in 
about the same time.\footnote{On a modern 6-core MacBook Pro with 32Gb memory,
these loading times were counted \emph{excluding} proof checking.}%
\end{isamarkuptext}\isamarkuptrue%
\begin{isamarkuptext}%
\begin{wrapfigure}{r}{0.2\textwidth}
  \includegraphics[width=0.2\textwidth]{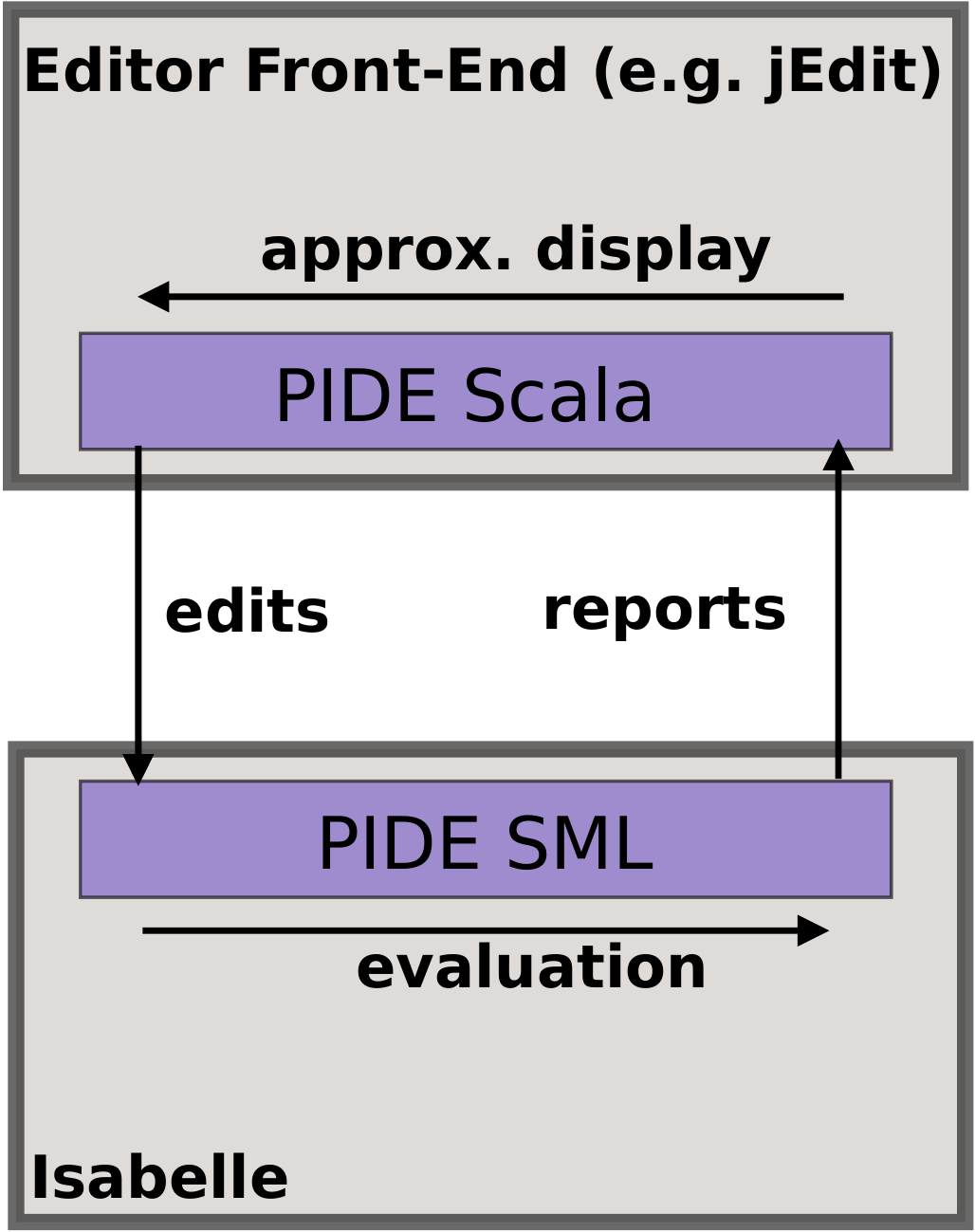}
  \vspace{-20pt}
\end{wrapfigure}
The PIDE (prover IDE) layer consists of a part written in SML
and another in Scala. Roughly speaking, PIDE implements ``continuous build and continuous check''
functionality over a textual albeit generic document model. It transforms user modifications of text
elements in an instance of this model into increments --- \emph{edits} --- and
communicates them to the Isabelle system. The latter reacts by the creation of a multitude of
light-weight reevaluation threads resulting in an asynchronous stream of
\emph{reports} containing \emph{markup} that is used to annotate
text elements in the editor front-end. For example, such markup is used to highlight variables or
keywords with specific colours, to hyperlink bound variables to their defining occurrences, or to
annotate type information to terms which become displayed by specific user gestures on demand (such
as hovering). Note that PIDE is not an editor, it is the framework that coordinates these
asynchronous information streams and optimizes their evaluation to a certain extent: outdated markup
referring to modified text is dropped, and corresponding re-calculations are oriented to the user
focus, for example. For PIDE, several editor applications have been developed, where Isabelle/jEdit
({\footnotesize \url{https://www.jedit.org}}) is the most commonly known. More
experimental alternatives based on Eclipse or Visual Studio Code exist.%
\end{isamarkuptext}\isamarkuptrue%
\isadelimdocument
\endisadelimdocument
\isatagdocument
\isamarkupsubsection{The PIDE Document Model%
}
\isamarkuptrue%
\endisatagdocument
{\isafolddocument}%
\isadelimdocument
\endisadelimdocument
\begin{isamarkuptext}%
\begin{wrapfigure}{r}{0.24\textwidth}
  \vspace{-12pt}
  \includegraphics[width=0.24\textwidth]{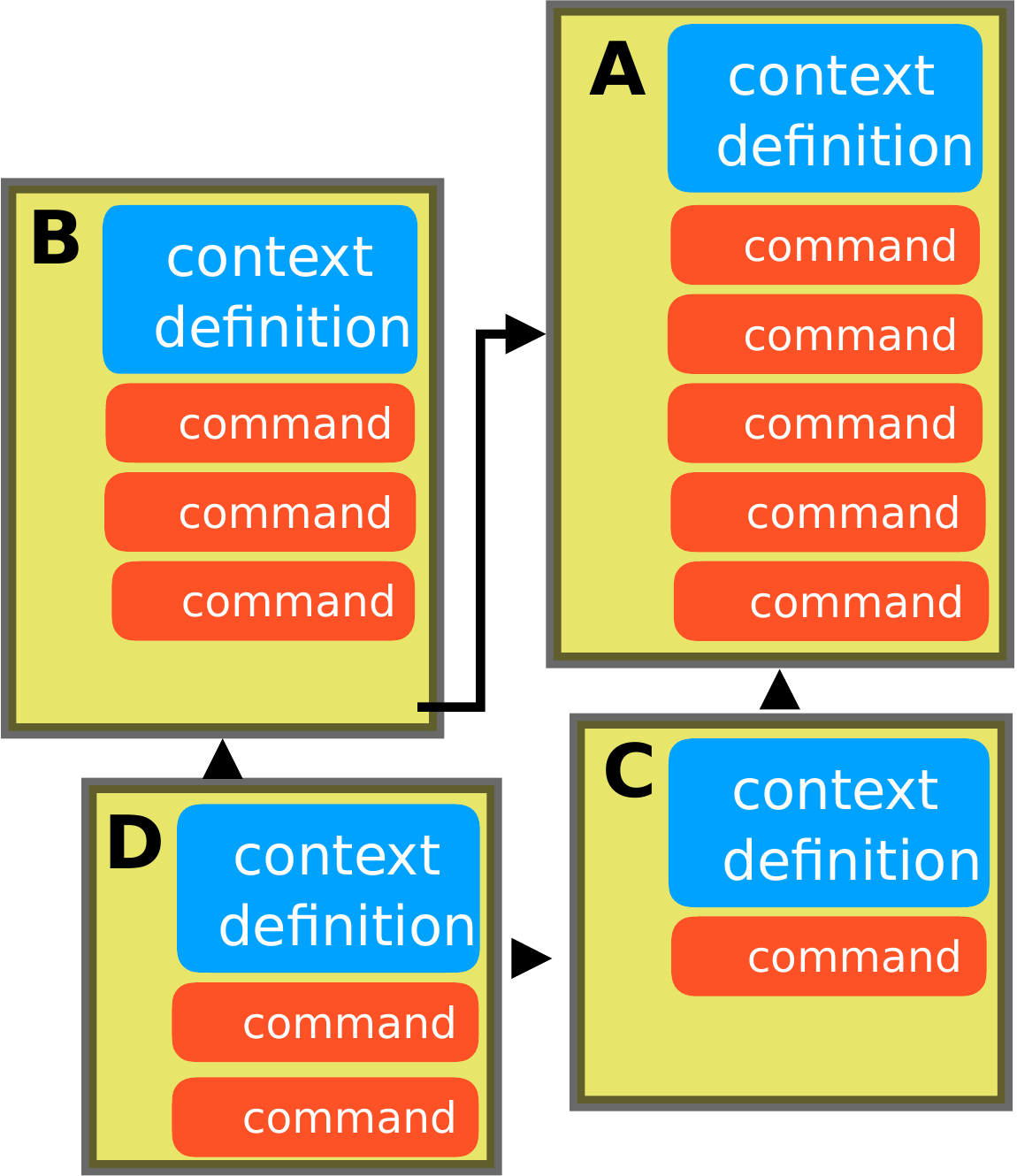}
  \vspace{-18pt}
\end{wrapfigure}
The document model foresees a number of atomic sub-documents (files), which are organized in the
form of an acyclic graph.
\begin{wrapfigure}{r}{0.33\textwidth}
\begin{isar}
theory C_Command
  imports C_Eval
  keywords "C" :: thy_decl
       and "C_file" :: thy_load
\end{isar}
\end{wrapfigure}
Such graphs can be grouped into sub-graphs called \emph{sessions} which can be compiled to binaries in 
order to avoid long compilation times --- Isabelle/C as such is a session. Sub-documents have a 
unique  name (the mapping to file paths in an underlying file-system is done in an integrated build 
management). The primary format of atomic sub-documents is \isatt{.thy} (historically for
``theory''), secondary formats can be \isatt{.sty}, \isatt{.tex}, \isatt{.c} or other sub-documents processed 
by Isabelle and listed in a configuration of the build system. 

A \isatt{.thy} file consists of a \emph{context
definition} and a body consisting of a sequence of \emph{commands}. The
context definition includes the sections \isa{\isakeyword{imports}} and
\isa{\isakeyword{keywords}}.  For example our context definition states that
\isa{C{\isacharunderscore}Command} is the name of the sub-document depending on
\isa{C{\isacharunderscore}Eval} which transitively includes the parser sources as (ML files)
sub-documents, as well as the C environment and the infrastructure for defining C level
annotations. \emph{Keywords} like \isa{\isacommand{C}} or
\isa{\isacommand{C{\isacharunderscore}file}} must be declared before use.

For this work, it is vital that predefined commands allow for the dynamic creation of
\emph{user-defined} commands similarly to the definition of new functions in a shell
interpreter. Semantically, commands are transition functions $\sigma \rightarrow \sigma$ where
$\sigma$ represents the system state called \emph{logical context}. The logical context in 
interactive provers contains --- among many other things --- the declarations of types, constant
symbols as well as the database with the definitions and established theorems.
A command starts with a pre-declared keyword followed by the specific syntax of this command; an
\emph{evaluation} of a command parses the input till the next command, and transfers
the parsed input to a transition function, which can be configured in a late binding table. Thus,
the evaluation of the generic document model allows for user programmed extensions including IDE and
document generation. 

Note that the Isabelle platform supports multiple syntax embeddings, i.e. the possibility
of nesting different language syntaxes inside the upper command syntax, using the
\isa{{\isacartoucheopen}\ {\isachardot}{\isachardot}\ {\isacartoucheclose}} brackets (such parsing techniques will be exploited in \csname isaDof.ref\endcsname[type={text}]{annotations}).  Accordingly, these syntactic sub-contexts may be
nested. In particular, in most of these sub-contexts, there may be a kind of semantic macro ---
called antiquotation and syntactically denoted in the format
\inlineisar+\at{name \<Open> .. \<Close>}+ --- that has access to the underlying logical context.
Similar to commands, user-defined antiquotations may be registered in a late-binding table. For
example, the standard \emph{term}-antiquotation in
\isa{\isacommand{ML}}\inlineisar|\<Open> val t = \at{term "3 +"} \<Close>| parses the argument \inlineisar|"3 +"| with
the Isabelle/HOL term parser, attempts to construct a $\lambda$-term in the internal
term representation and to bind it to \isatt{t}; however, this fails (the plus operation is
declared infix in logical context) and therefore the entire command fails.%
\end{isamarkuptext}\isamarkuptrue%
\isadelimdocument
\endisadelimdocument
\isatagdocument
\isamarkupsubsection{Some Basics of PIDE Programming%
}
\isamarkuptrue%
\endisatagdocument
{\isafolddocument}%
\isadelimdocument
\endisadelimdocument
\begin{isamarkuptext}%
\begin{wrapfigure}{r}{0.47\textwidth}
  \vspace{-5pt}
\begin{isar}
ML\<Open> val pos = \at{here};
    val markup = Position.here pos;
    writeln ("And a link to the declaration\
              \ of 'here' is " ^ markup) \<Close>
\end{isar}
  \vspace{-15pt}
\end{wrapfigure}
A basic data-structure relevant for PIDE is \emph{positions}; beyond the usual line
and column information they can represent ranges, list of continuous ranges, and the name of the
atomic sub-document in which they are contained. It is straightforward to
use the antiquotation \inlineisar|\at{here}| to infer from the system lexer the actual position of
the antiquotation in the global document. The system converts the position to a markup
representation (a string representation) and sends the result via \isatt{writeln} to
the interface.%
\end{isamarkuptext}\isamarkuptrue%
\begin{isamarkuptext}%
\begin{wrapfigure}{r}{0.4\textwidth}
  \vspace{-8pt}
  \includegraphics[width=0.4\textwidth]{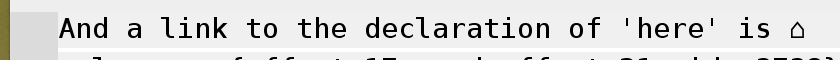}
  \vspace{-20pt}
\end{wrapfigure}
In return, the PIDE output window shows the little house-like
symbol $\house$, which is actually hyperlinked to the position of \inlineisar|\at{here}|. The ML
structures \isatt{Markup} and \isatt{Properties}
represent the basic libraries for annotation data which is part of the protocol sent from Isabelle
to the front-end. They are qualified as ``quasi-abstract'', which means they are intended to be an
abstraction of the serialized, textual presentation of the protocol. A markup must be tagged with a
unique id; this is done by the library \isatt{serial} function. Typical code for taking
a string \isatt{cid} from the editing window, together with its position
\isatt{pos}, and sending a specific markup referring to this in the editing window
managed by PIDE looks like this:
\begin{isar}
ML\<Open> fun report_def_occur pos cid = Position.report pos (my_markup true cid (serial ()) pos) \<Close>
\end{isar}
Note that \isatt{my{\char`\_}markup} (not shown here) generates the layout attributes of the link and 
that the \isatt{true} flag is used for markup declaring \isatt{cid} as a defining occurrence, i.e. 
as \emph{target} (rather than the \emph{source}) in the hyperlink animation in PIDE.%
\end{isamarkuptext}\isamarkuptrue%
\begin{isamarkupsection*}%
[label = {frontend_arch},type = {scholarly_paper.technical}, args={label = {frontend_arch},type = {scholarly_paper.technical}, Isa_COL.text_element.level = {}, Isa_COL.text_element.referentiable = {False}, Isa_COL.text_element.variants = {{STR ''outline'', STR ''document''}}, scholarly_paper.text_section.main_author = {}, scholarly_paper.text_section.fixme_list = {}, Isa_COL.text_element.level = {}, scholarly_paper.technical.definition_list = {}}]The C11 Parser Generation Process and Architecture%
\end{isamarkupsection*}\isamarkuptrue%
\begin{isamarkuptext}%
Isabelle uses basically two parsing technologies:

\begin{enumerate}%
\item Earley parsing \cite{DBLP:journals/cacm/Earley70}
intensively used for mixfix-syntax denoting $\lambda$-terms in mathematical notation,

\item combinator parsing \cite{DBLP:journals/jfp/Hutton92} typically used for high-level 
command syntax.%
\end{enumerate}%
\end{isamarkuptext}\isamarkuptrue%
\begin{isamarkuptext}%
Both technologies offer the dynamic extensibility necessary for Isabelle as an
interactive platform geared towards incremental development and sophisticated mathematical
notations. However, since it is our goal to support \emph{programming languages} in
a fast parse-check-eval cycle inside an IDE, we opt for a Lex and Yacc deterministic grammar
approach. It turns out the resulting automata based parser performs well enough for our purpose; the
gain in performance is discussed in the next section.%
\end{isamarkuptext}\isamarkuptrue%
\begin{isamarkuptext}%
In the following, we describe a novel technique for the construction and integration of
this type of parser into the Isabelle platform. Since it is mostly relevant for integrators copying
our process to similar languages such as JavaScript or Rust \footnote{E.g. {\footnotesize
  \url{http://hackage.haskell.org/package/language-javascript}} or {\footnotesize
  \url{http://hackage.haskell.org/package/language-rust}}}, users of the
Isabelle/C platform may skip this section: for them, the take-home message is that the overall
generation process takes about 1 hour, the compilation of the generated files takes 15s, and that
the generated files should be fairly portable to future Isabelle versions.%
\end{isamarkuptext}\isamarkuptrue%
\begin{isamarkuptext}%
We base our work on the C11 parsing library {\footnotesize
  \url{http://hackage.haskell.org/package/language-c}} implemented in Haskell by
Huber, Chakravarty, Coutts and Felgenhauer; we particularly focus on its open-source Haskell Yacc
grammar as our starting point. We would like to emphasize that this is somewhat arbitrary, our build
process can be easily adapted to more recent versions when available.%
\end{isamarkuptext}\isamarkuptrue%
\begin{isamarkupfigure*}%
[label = {architecture},type = {Isa_COL.figure}, args={label = {architecture},type = {Isa_COL.figure}, Isa_COL.figure.relative_width = {70}, Isa_COL.figure.src = {figures/C11-Package-Architecture}, Isa_COL.figure.spawn_columns = {True}}]The architecture of Isabelle/C%
\end{isamarkupfigure*}\isamarkuptrue%
\begin{isamarkuptext}%
The diagram in \csname isaDof.ref\endcsname[type={Isa_COL.figure}]{architecture} presents the architecture of
Isabelle/C. The original Haskell library was not modified, it is presented in blue together with
generated sources, in particular the final two blue boxes represent about 11~kLoC. In output, the
glue code in brown constitutes the core implementation of Isabelle/C, amounting to 6~kLoC (without
yet considering semantic back-ends).%
\end{isamarkuptext}\isamarkuptrue%
\begin{isamarkupsubsection*}%
[label = {arch1},type = {scholarly_paper.technical}, args={label = {arch1},type = {scholarly_paper.technical}, Isa_COL.text_element.level = {}, Isa_COL.text_element.referentiable = {False}, Isa_COL.text_element.variants = {{STR ''outline'', STR ''document''}}, scholarly_paper.text_section.main_author = {}, scholarly_paper.text_section.fixme_list = {}, Isa_COL.text_element.level = {}, scholarly_paper.technical.definition_list = {}}]Generating the AST%
\end{isamarkupsubsection*}\isamarkuptrue%
\begin{isamarkuptext}%
In the following, we refer to \emph{languages} by $\mathcal{L}$,
$\mathcal{I}$. The notation $\text{AST}^{\mathcal{L}}_{\mathcal{I}}$ refers to abstract syntaxes for
language $\mathcal{L}$ implemented in language $\mathcal{I}$. For example, we refer by
$\text{AST}^{\text{C11}}_{\text{ML}}$ to an AST implementation of C11 implemented in SML. Indices
will be dropped when no confusion arises, or to highlight the fact that our approach is
sufficiently generic.%
\end{isamarkuptext}\isamarkuptrue%
\begin{isamarkuptext}%
For our case, we exploit that from a given Haskell source $\text{AST}_{\text{HS}}$,
Haskabelle generates to a maximum extent an Isabelle/HOL theory. Via the Isabelle code generator, an
$\text{AST}_{\text{ML}}$ can be obtained from a constructive $\text{AST}_{\text{HOL}}$
representation. However, the process is challenging for technical reasons in practice due to the
enormous size of $\text{AST}^{\text{C11}}$ (several hundreds of constructors), and due to certain
type declarations not initially supported by Haskabelle (we have to implement here the necessary
features). Ultimately, the process to compile $\text{AST}_{\text{HS}}$ to $\text{AST}_{\text{ML}}$
is done only once at build time, it comprises:

\begin{enumerate}%
\item the generation of $\text{AST}_{\text{HOL}}$ from $\text{AST}_{\text{HS}}$, represented as a
collection of \isa{\isacommand{datatype}},

\item the execution of the \isa{\isacommand{datatype}} theory for $\text{AST}_{\text{HOL}}$ and checking of all their
proofs,\footnote{Large mutually recursive datatypes in $\text{AST}_{\text{HOL}}$ might lead to worse 
performance time,
see for instance
  {\footnotesize \url{https://lists.cam.ac.uk/pipermail/cl-isabelle-users/2016-March/msg00034.html}}
  and
  {\footnotesize \url{https://lists.cam.ac.uk/pipermail/cl-isabelle-users/2017-April/msg00000.html}}.}

\item the generation of an $\text{AST}_{\text{ML}}$ from $\text{AST}_{\text{HOL}}$.%
\end{enumerate}%
\end{isamarkuptext}\isamarkuptrue%
\begin{isamarkupsubsection*}%
[label = {lexer},type = {scholarly_paper.technical}, args={label = {lexer},type = {scholarly_paper.technical}, Isa_COL.text_element.level = {}, Isa_COL.text_element.referentiable = {False}, Isa_COL.text_element.variants = {{STR ''outline'', STR ''document''}}, scholarly_paper.text_section.main_author = {}, scholarly_paper.text_section.fixme_list = {}, Isa_COL.text_element.level = {}, scholarly_paper.technical.definition_list = {}}]Constructing a Lexer for C11%
\end{isamarkupsubsection*}\isamarkuptrue%
\begin{isamarkuptext}%
We decided against the option of importing the equivalent Haskell lexer, as it is
coming under-developed compared to the existing PIDE lexer library, natively
supporting Unicode-like symbols (mostly for annotations). Using a more expressive position
data-structure, our C lexer is also compatible with the native ML lexer regarding the handling of
errors and backtracking (hence the perfect fit when nesting one language inside the other). Overall,
the modifications essentially boil down to taking an extreme care of comments and directives which
have intricate lexical conventions (see \csname isaDof.ref\endcsname[type={text}]{lexer_ex}).%
\end{isamarkuptext}\isamarkuptrue%
\begin{isamarkupsubsection*}%
[label = {parser},type = {scholarly_paper.technical}, args={label = {parser},type = {scholarly_paper.technical}, Isa_COL.text_element.level = {}, Isa_COL.text_element.referentiable = {False}, Isa_COL.text_element.variants = {{STR ''outline'', STR ''document''}}, scholarly_paper.text_section.main_author = {}, scholarly_paper.text_section.fixme_list = {}, Isa_COL.text_element.level = {}, scholarly_paper.technical.definition_list = {}}]Generating the Shift-Reduce Parser from the Grammar%
\end{isamarkupsubsection*}\isamarkuptrue%
\begin{isamarkuptext}%
In the original C11 library, together with $\text{AST}_{\text{HS}}$, there is a Yacc
grammar file $\text{G}_{\text{HS-YACC}}$ included, which we intend to use to conduct the C
parsing. However due to technical limitations of Haskabelle (and advanced Haskell constructs in the
associated $\text{G}_{\text{HS}}$), we do not follow the same approach as \csname isaDof.ref\endcsname[type={text}]{arch1}. Instead, an ultimate grammar $\text{G}_{\text{ML}}$ is obtained by letting
ML-Yacc participate in the generation process. In a nutshell, the overall grammar translation chain
becomes: $\text{G}_{\text{HS-YACC}} \longrightarrow_{\text{HS}} \text{G}_{\text{ML-YACC}}
\longrightarrow_{\text{ML}} \text{G}_{\text{ML}}$.

$\longrightarrow_{\text{HS}}$ is implemented by modifying the Haskell parser generator Happy,
because Happy is already natively supporting the whole $\mathcal{L}_{\text{HS-YACC}}$. Due to the
close connection between Happy and ML-Yacc, the translation is even almost linear. However cares
must be taken while translating monadic rules~\footnote{\url{https://www.haskell.org/happy/doc/html/sec-monads.html}} of
$\text{G}_{\text{HS-YACC}}$, as $\mathcal{L}_{\text{ML-YACC}}$ does not support such rules. In
$\text{G}^{\text{C11}}$, monadic rules are particularly important for scoping analyses, or while
building new informative AST nodes (in contrast to disambiguating non-monadic rules, see
\isa{{\isacharat}} vs. \isa{{\isacharampersand}} in \csname isaDof.ref\endcsname[type={text}]{annotations}).
Consequently, applying ML-Yacc $\longrightarrow_{\text{ML}}$ on $\text{G}_{\text{ML-YACC}}$ is not
enough: after compiling $\text{G}_{\text{ML}}$ to an efficient Shift-Reduce automaton, we
substantially modified the own grammar interpreter of ML-Yacc to implement all features of
$\mathcal{L}_{\text{HS-YACC}}$ presented as used in $\text{G}_{\text{HS-YACC}}$.%
\end{isamarkuptext}\isamarkuptrue%
\begin{isamarkupsection*}%
[label = {ctests},type = {scholarly_paper.technical}, args={label = {ctests},type = {scholarly_paper.technical}, Isa_COL.text_element.level = {}, Isa_COL.text_element.referentiable = {False}, Isa_COL.text_element.variants = {{STR ''outline'', STR ''document''}}, scholarly_paper.text_section.main_author = {}, scholarly_paper.text_section.fixme_list = {}, Isa_COL.text_element.level = {}, scholarly_paper.technical.definition_list = {}}]Isabelle/C: Syntax Tests and Experimental Results%
\end{isamarkupsection*}\isamarkuptrue%
\begin{isamarkuptext}%
The question arises, to what extent our construction provides a faithful parser for
C11, and if Isabelle/C is sufficiently stable and robust to handle real world sources. A related
question is the treatment of \isatt{cpp} preprocessing directives: while a
minimal definition of the preprocessor is part of C standards since C99, practical implementations
vary substantially.  Moreover, \isatt{cpp} comes close to be Turing complete:
recursive computations can be specified, but the expansion strategy bounds the number of
unfolding. Therefore, a complete \isatt{cpp} reimplementation contradicts our
objective to provide efficient IDE support inside Isabelle. Instead, we restrict ourselves to a
common subset of macro expansions and encourage, whenever possible, Isabelle specific mechanisms
such as user programmed C annotations. C sources depending critically on a specific
\isatt{cpp} will have to be processed outside
Isabelle.\footnote{Isabelle/C has a particular option to activate (or not) an automated
call to \isatt{cpp} before any in-depth treatment.}%
\end{isamarkuptext}\isamarkuptrue%
\begin{isamarkupsubsection*}%
[label = {lexer_ex},type = {scholarly_paper.technical}, args={label = {lexer_ex},type = {scholarly_paper.technical}, Isa_COL.text_element.level = {}, Isa_COL.text_element.referentiable = {False}, Isa_COL.text_element.variants = {{STR ''outline'', STR ''document''}}, scholarly_paper.text_section.main_author = {}, scholarly_paper.text_section.fixme_list = {}, Isa_COL.text_element.level = {}, scholarly_paper.technical.definition_list = {}}]Preprocessing Lexical Conventions: Comments and Newlines%
\end{isamarkupsubsection*}\isamarkuptrue%
\begin{isamarkuptext}%
A very basic standard example taken from the GCC / CPP documentation~\footnote{\url{https://gcc.gnu.org/onlinedocs/cpp/Initial-processing.html}}
shows the quite intricate mixing of comment styles that
represents a challenge for our C lexer. A further complication is that it is allowed and common
practice to use backslash-newlines \isa{{\isacharbackslash}{\isasymnewline}} \emph{anywhere} in C
sources, be it inside comments, string denotations, or even regular C keywords like
\isa{i{\isacharbackslash}{\isasymnewline}n{\isacharbackslash}{\isasymnewline}t} (see also \autoref{fig:clean}).

\begin{wrapfigure}{r}{0.42\textwidth}
  \vspace{-12pt}
  \includegraphics[width=0.42\textwidth]{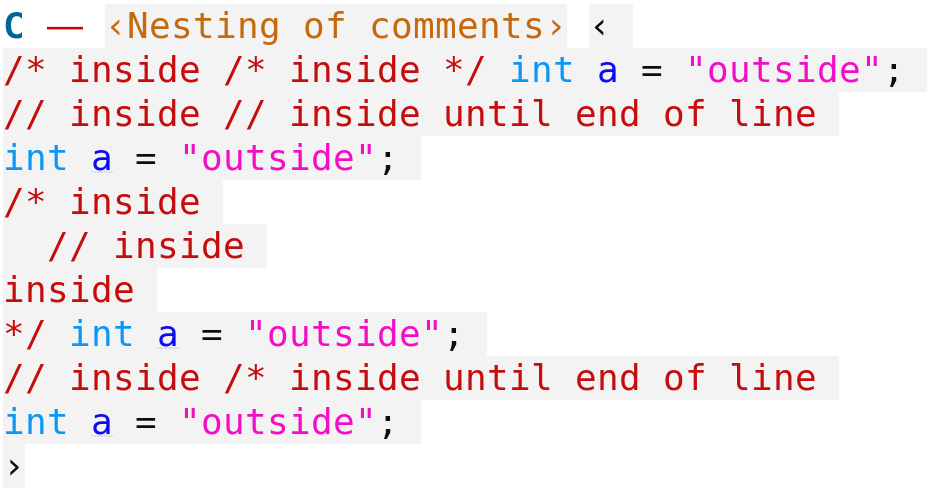}
  \vspace{-20pt}
\end{wrapfigure}
In fact, many C processing tools assume that all comments have already been removed via
\isatt{cpp} before they start any processing. However, annotations in comments carry relevant information 
for back-ends as shown in  \csname isaDof.ref\endcsname[type={text}]{annotations}. Consequently, they must be explicitly 
represented in $\text{AST}^{\text{C11}}_{\text{ML}}$,
whereas the initial $\text{AST}^{\text{C11}}_{\text{HS}}$ 
is not designed to carry such extra information. 
Annotations inside comments may again contain structured information like programming code, formulas,
and proofs, which implies the need for nested syntax. Fortunately, Isabelle is designed to
manage multiple parsing layers with the technique of \emph{cascade sources}~\footnote{\url{http://isabelle.in.tum.de/repos/isabelle/file/83774d669b51/src/Pure/General/source.ML}}
(see also \csname isaDof.ref\endcsname[type={text}]{C-sample3}). We exploit this infrastructure to integrate
back-end specific syntax and annotation semantics based on the parsing technologies available.%
\end{isamarkuptext}\isamarkuptrue%
\isadelimdocument
\endisadelimdocument
\isatagdocument
\isamarkupsubsection{Preprocessing Side-Effects: Antiquoting Directives vs. Pure Annotations%
}
\isamarkuptrue%
\endisatagdocument
{\isafolddocument}%
\isadelimdocument
\endisadelimdocument
\begin{isamarkuptext}%
Whereas \emph{comments} can be safely removed without affecting the meaning of C code,
\emph{directives} are semantically relevant for compilation and evaluation.

\begin{enumerate}%
\item Classical directives: \isa{{\isacharhash}define\ x\ TOKS} makes any incoming C identifier
\isa{x} be replaced by some \emph{arbitrary} tokens \isa{TOKS},
even when included via the \isa{{\isacharhash}include} directive.

\item Typed (pseudo-)directives as commands: It is easy to overload or implement a new
\isa{{\isacharhash}define{\isacharprime}} acting only on a decided subset of well-formed \isa{TOKS}. There
are actually no differences between Isabelle/C directives and Isabelle commands: both are
internally of type $\sigma \rightarrow \sigma$ (see \csname isaDof.ref\endcsname[type={text}]{background}).

\item Non-expanding annotations: Isabelle/C annotations
$\isa{{\isacharslash}{\isacharasterisk}{\isacharat}}~\mathcal{L}_\text{annot}~\isa{{\isacharasterisk}{\isacharslash}}$ or
$\isa{{\isacharslash}{\isacharslash}{\isacharat}}~\mathcal{L}_\text{annot}$ can be freely intertwined between other tokens, even
inside directives. In contrast to (antiquoting) directives and similarly as C comments, their
designed intent is to not modify the surrounding parsing code.%
\end{enumerate}%
\end{isamarkuptext}\isamarkuptrue%
\begin{isamarkuptext}%
A limitation of Isabelle and its current document model is that there is no way for user
programmed extensions to exploit implicit dependencies between
sub-documents. Thus, a sub-document referred to via \isa{{\isacharhash}include\ {\isacharless}some{\isacharunderscore}file{\isachargreater}} will not
lead to a reevaluation of a \isa{\isacommand{C}{\isacartoucheopen}\ {\isachardot}{\isachardot}\ {\isacartoucheclose}} command whenever modified. (The only 
workaround is to open all transitively required sub-documents  \emph{by hand}.)%
\end{isamarkuptext}\isamarkuptrue%
\begin{isamarkupsubsection*}%
[label = {parser_validation},type = {scholarly_paper.technical}, args={label = {parser_validation},type = {scholarly_paper.technical}, Isa_COL.text_element.level = {}, Isa_COL.text_element.referentiable = {False}, Isa_COL.text_element.variants = {{STR ''outline'', STR ''document''}}, scholarly_paper.text_section.main_author = {}, scholarly_paper.text_section.fixme_list = {}, Isa_COL.text_element.level = {}, scholarly_paper.technical.definition_list = {}}]A Validation via the seL4 Test Suite%
\end{isamarkupsubsection*}\isamarkuptrue%
\begin{isamarkuptext}%
The AutoCorres environment contains a C99 parser developed by Michael Norrish \cite{DBLP:conf/sosp/KleinEHACDEEKNSTW09}. Besides a parser test-suite, there is the entire seL4
codebase (written in C99) which has been used for the code verification part of the seL4
project. While the parser in itself represents a component belonging to the trusted base of the
environment, it is arguably the most tested parser for a semantically well-understood translation in
a proof environment today.%
\end{isamarkuptext}\isamarkuptrue%
\begin{isamarkuptext}%
It is therefore a valuable reference for a comparison test, especially since
$\text{AST}^\text{C99}$ and $\text{AST}^\text{C11}$ are available in the same implementation
language. From $\text{AST}^\text{C11}_{\text{HOL}}$ to $\text{AST}^\text{C99}_{\text{HOL}}$ we
construct an abstraction function $C^\downarrow$. A detailed description of
$C^\downarrow$ is out of the scope of this paper; we would like to mention that it was 4 man-months
of work due to the richness of $\text{AST}^\text{C11}$.
As such, the abstraction function $C^\downarrow$ is at the heart of the AutoCorres
integration into our framework described in \csname isaDof.ref\endcsname[type={text}]{autocorres}. Note that
$\text{AST}^{\text{C99}}$ seems to be already an abstraction compared to the C99 standard. This
gives rise to a particular testing methodology: we can compile the test suites as well as the seL4
source files by both ML parsers $\text{PARSE}_{\text{\emph{stop}}}^{\text{C99}}$ and
$\text{PARSE}_{\text{\emph{report}}}^{\text{C11}}$, abstract the output of the latter via
$C^\downarrow$ and compare the results.%
\end{isamarkuptext}\isamarkuptrue%
\begin{isamarkuptext}%
Our test establishes that both parsers agree on the entire seL4 codebase. However
trying to compare the two parsers using other criteria is not possible, for example we had to limit
ourselves to C programs written in a subset of C99. Fundamentally, the two parsers are achieving
different tasks: the one of $\text{PARSE}_{\text{\emph{stop}}}$ is to just return a parsed AST. In
contrast, $\text{PARSE}_{\text{\emph{report}}}$ intends to maximize markup reporting, irrespective
of a final parsing success or failure, and reports are provided in parallel during its (monadic)
parsing activity. Thus, in the former scenario, the full micro-kernel written in 26~kLoC can be
parsed in 0.1s. In the latter, all reports we have thought helpful to implement are totally rendered
before 20s. Applying $C^\downarrow$ takes 0.02 seconds, so our $\text{PARSE}_{\text{\emph{report}}}$
gives an average of 2s for a 2-3~kLoC source. By interweaving a source with proofs referring to the
code elements, the responsiveness of PIDE should therefore be largely sufficient.%
\end{isamarkuptext}\isamarkuptrue%
\begin{isamarkupsection*}%
[label = {annotations},type = {scholarly_paper.technical}, args={label = {annotations},type = {scholarly_paper.technical}, Isa_COL.text_element.level = {}, Isa_COL.text_element.referentiable = {False}, Isa_COL.text_element.variants = {{STR ''outline'', STR ''document''}}, scholarly_paper.text_section.main_author = {}, scholarly_paper.text_section.fixme_list = {}, Isa_COL.text_element.level = {}, scholarly_paper.technical.definition_list = {}}]Generic Semantic Annotations for C%
\end{isamarkupsection*}\isamarkuptrue%
\begin{isamarkuptext}%
With respect to interaction with the underlying proof-engine, there are essentially two
lines of thought in the field of deductive verification techniques:

\begin{enumerate}%
\item either programs and specifications --- i.e. the pre- and post-condition contracts --- are clearly 
separated, or

\item the program is annotated with the specification, typically by using some form of formal
comment.%
\end{enumerate}%
\end{isamarkuptext}\isamarkuptrue%
\begin{isamarkuptext}%
Of course, it is possible to inject the essence of annotated specifications directly
into proofs, e.g. by instantiating the \isa{while} rule of the Hoare calculus by the needed
invariant inside the proof script. The resulting clear separation of programs from proofs may be
required by organisational structures in development projects. However, in many cases, modelling
information may be interesting for programmers, too. Thus, having pre- and post-conditions locally
in the source close to its point of relevance increases its maintainability. It became therefore
common practice to design languages with annotations, i.e. structured comments
\emph{inside} a programming source. Examples are ACSL standardized by ANSI/ISO (see
  {\footnotesize \url{https://frama-c.com/download/acsl.pdf}}) or UML/OCL \cite{DBLP:journals/afp/BruckerTW14} for static analysis tools. Isabelle/C supports both the
inject-into-proof style and annotate-the-source style in its document model; while the former is
kind of the default, we address in this section the necessary technical infrastructure for the
latter.%
\end{isamarkuptext}\isamarkuptrue%
\begin{isamarkuptext}%
Generally speaking, a generic annotation mechanism which is sufficiently expressive to
capture idioms used in, e.g., Frama-C, Why3, or VCC is more problematic than one might think.
Consider this:
\begin{isar}
     for (int i = 0; i < n; i++) a+= a*i /*@ annotation */
\end{isar}
To which part of the AST does the annotation refer? To \isa{i}? \isa{a{\isacharasterisk}i}? The
assignment? The loop? Some verification tools use prefix annotations (as in Why3 for procedure
contracts), others even a kind of parenthesis of the form:
\begin{isar}
     /*@ annotation_begin */ ...  /*@ annotation_end */
\end{isar}
The matter gets harder since the C environment --- a table mapping C identifiers to their type and
status --- changes according to the reference point in the AST. This means that the context relevant
to type-check an annotation such as \isa{{\isacharslash}{\isacharasterisk}{\isacharat}\ assert\ {\isasymopen}a\ {\isachargreater}\ i{\isasymclose}\ {\isacharasterisk}{\isacharslash}} strongly
differs depending on the annotation's position. And the matter gets even further complicated since
Isabelle/C lives inside a proof environment; here, local theory development (rather than bold ad-hoc
axiomatizations) is a major concern.

\begin{wrapfigure}{r}{0.56\textwidth}
  \vspace{-10pt}
  \includegraphics[width=0.56\textwidth]{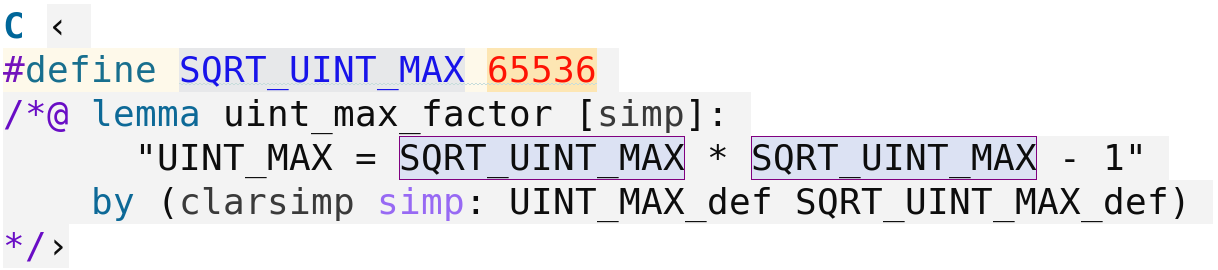}
  \vspace{-20pt}
\end{wrapfigure}
The desire for fast impact analysis resulting from changes may inspire one to annotate local proofs
near directives, which is actually what is implemented in our Isabelle/C/AutoCorres example
(\csname isaDof.ref\endcsname[type={text}]{backends}). In the example, the semantic back-end converts the
\isatt{cpp} macro into a HOL \emph{definition}, i.e. an extension
of the underlying theory context by the conservative axiom
\isa{SQRT{\isacharunderscore}UINT{\isacharunderscore}MAX\ {\isasymequiv}\ {\isadigit{6}}{\isadigit{5}}{\isadigit{5}}{\isadigit{3}}{\isadigit{6}}} bound to the name
\isa{SQRT{\isacharunderscore}UINT{\isacharunderscore}MAX{\isacharunderscore}def}. This information is used in the subsequent proof
establishing a new theory context containing the lemma \isa{uint{\isacharunderscore}max{\isacharunderscore}factor}
configured to be used as rewrite rule whenever possible in future proofs. This local lemma
establishes a relation of \isa{SQRT{\isacharunderscore}UINT{\isacharunderscore}MAX} to the maximally representable number
\isa{UINT{\isacharunderscore}MAX} for an unsigned integer according to the underlying memory model.

Obviously, the scheduling of these transformations of the underlying theory contexts is
non-trivial.%
\end{isamarkuptext}\isamarkuptrue%
\begin{isamarkupfigure*}%
[label = {C-sample3},type = {Isa_COL.figure}, args={label = {C-sample3},type = {Isa_COL.figure}, Isa_COL.figure.relative_width = {100}, Isa_COL.figure.src = {figures/A-C-Source3}, Isa_COL.figure.spawn_columns = {True}}]Advanced annotation programming%
\end{isamarkupfigure*}\isamarkuptrue%
\begin{isamarkupsubsection*}%
[label = {navigation},type = {scholarly_paper.technical}, args={label = {navigation},type = {scholarly_paper.technical}, Isa_COL.text_element.level = {}, Isa_COL.text_element.referentiable = {False}, Isa_COL.text_element.variants = {{STR ''outline'', STR ''document''}}, scholarly_paper.text_section.main_author = {}, scholarly_paper.text_section.fixme_list = {}, Isa_COL.text_element.level = {}, scholarly_paper.technical.definition_list = {}}]Navigation for Annotation Commands%
\end{isamarkupsubsection*}\isamarkuptrue%
\begin{isamarkuptext}%
In order to overcome the problem of syntactic ambiguity of annotations, we slightly
refine the syntax of semantic annotations by the concept of a navigation expression:
\begin{isar}
    $\mathcal{L}_\text{annot}$ $=$ $\varnothing$ | <navigation-expr> <annotation-command> $\mathcal{L}_\text{annot}$ 
\end{isar}

A {\footnotesize\texttt{<navigation-expr>}} string consists of a sequence of \isa{{\isacharplus}} symbols followed
by a sequence consisting of \isa{{\isacharat}} or \isa{{\isacharampersand}} symbols. It allows for navigating
in the syntactic context, by advancing tokens with several \isa{{\isacharplus}}, or taking an ancestor
AST node with several \isa{{\isacharat}} (or \isa{{\isacharampersand}} which only targets monadic grammar rules). This
corresponds to a combination of right-movements in the AST, and respectively parent-movements. This
way, the ``focus'' of an {\footnotesize\texttt{<annotation-command>}} can be modified to denote any C
fragment of interest.%
\end{isamarkuptext}\isamarkuptrue%
\begin{isamarkuptext}%
As a relevant example for debugging, consider \csname isaDof.ref\endcsname[type={Isa_COL.figure}]{C-sample3}.
The annotation command \isa{\isacommand{highlight}} is a predefined Isabelle/C ML-library function that is
interpreted as C annotation. Its code is implicitly parameterized by the syntactical context,
represented by \isatt{stack{\char`\_}top} whose type is a subset of
$\text{AST}^\text{C11}$, and the lexical environment \isatt{env} containing the
lexical class of identifiers, scopes, positions and serials for markup. The navigation string before
\isa{\isacommand{highlight}} particularly influences which
\isatt{stack{\char`\_}top} value gets ultimately selected. The third screenshot in \csname isaDof.ref\endcsname[type={Isa_COL.figure}]{C-sample3} demonstrates the influence of the static environment: an Isabelle/C
predefined command \isa{\isacommand{{\isasymsimeq}setup}} allows for ``recursively'' calling the C
environment itself. This results in the export of definitions in the surrounding logical context,
where the propagation effect may be controlled with options like
\isa{C{\isacharunderscore}starting{\isacharunderscore}env}. \isa{\isacommand{{\isasymsimeq}setup}} actually
mimics standard Isabelle \isa{\isacommand{setup}} command, but extends it by
\isatt{stack{\char`\_}top} and \isatt{env} \footnote{cf. \url{https://isabelle.in.tum.de/doc/isar-ref.pdf}}. In the
example, the first recursive call uses \isatt{env} allowing it to detect that
\isa{b} is a local parameter, while the second ignores it which results in a treatment as a
free global variable. Note that bound global variables are not green but depicted in black.%
\end{isamarkuptext}\isamarkuptrue%
\begin{isamarkupsubsection*}%
[label = {annot1},type = {scholarly_paper.technical}, args={label = {annot1},type = {scholarly_paper.technical}, Isa_COL.text_element.level = {}, Isa_COL.text_element.referentiable = {False}, Isa_COL.text_element.variants = {{STR ''outline'', STR ''document''}}, scholarly_paper.text_section.main_author = {}, scholarly_paper.text_section.fixme_list = {}, Isa_COL.text_element.level = {}, scholarly_paper.technical.definition_list = {}}]Defining Annotation Commands%
\end{isamarkupsubsection*}\isamarkuptrue%
\begin{isamarkuptext}%
Extending the default configuration of commands, text and code antiquotations from the
Isabelle platform to Isabelle/C is straightforward. For example, the central Isabelle command
definition:
\begin{isar}
    Outer_Syntax.command: $K_\text{\emph{cmd}}$ -> ($\sigma$ -> $\sigma$) parser -> unit
\end{isar}
establishes the dynamic binding between a command keyword $K_\text{\emph{cmd}} =
\isa{\isacommand{definition}} \isatt{|}
\isa{\isacommand{lemma}} \isatt{|} \dots $ and a parser, whose value is
a system transition.\footnote{$\sigma$ has actually the internal Isabelle type
  \isatt{Toplevel.transition}.} The \isatt{parser} type
stems from the
aforementioned parser combinator library: \isatt{{\char`\'}a\ parser}
\isatt{=} \isatt{Token.T\ list\ {\char`\-}{\char`\>}\ {\char`\'}a\ *\ Token.T\ list}.%
\end{isamarkuptext}\isamarkuptrue%
\begin{isamarkuptext}%
Analogously, Isabelle/C provides an internal late-binding table for
\emph{annotation commands}:
\begin{isar}
    C_Annotation.command : $K_\text{\emph{cmd}}$ -> ($\text{<navigation-expr>}$ -> $R_\text{\emph{cmd}}$ c_parser) -> unit
    C_Annotation.command': $K_\text{\emph{cmd}}$ -> ($\text{<navigation-expr>}$ -> $R_\text{\emph{cmd}}$ c_parser) -> $\sigma$ -> $\sigma$
    C_Token.syntax': 'a parser -> 'a c_parser
\end{isar}
where in this paper we define $R_\text{\emph{cmd}} = \sigma~\isatt{{\char`\-}{\char`\>}}~\sigma$ as
above.\footnote{In some parallel work, we focus on running commands in native efficient
  speed with $R_\text{\emph{cmd}} = \isatt{(} K_\text{\emph{cmd}}
  \isatt{*} \isatt{(}\sigma \isatt{{\char`\-}{\char`\>}} \sigma
  \isatt{)} \isatt{)} \isatt{list}$. \cite{DBLP:journals/afp/TuongW15}}
Since the type \isatt{c{\char`\_}parser} is isomorphic to \isatt{parser},
but accepting C tokens, one can use \isatt{C{\char`\_}Token.syntax{\char`\'}} to translate and carry
the default Isar commands \emph{inside} the \isa{\isacommand{C}{\isacartoucheopen}\ {\isachardot}{\isachardot}\ {\isacartoucheclose}} scope, such as \isa{\isacommand{lemma}} or
\isa{\isacommand{by}}. Using \isa{\isacommand{{\isasymsimeq}setup}}, one can even
define an annotation command \isa{\isacommand{C}} taking a C code as argument, as the ML
code of \isa{\isacommand{{\isasymsimeq}setup}} has type
$\alpha^{\text{AST}}~\isatt{{\char`\-}{\char`\>}\ env\ {\char`\-}{\char`\>}}~R_\text{\emph{cmd}}$ (which is enough for
calling \isatt{C{\char`\_}Annotation.command{\char`\'}} in the ML code). Here, whereas the type
\isatt{env} is always the same, the type $\alpha^{\text{AST}} \subseteq
\text{AST}^\text{C11}$ varies depending on {\footnotesize\texttt{<navigation-expr>}} (see \csname isaDof.ref\endcsname[type={text}]{annot2}).%
\end{isamarkuptext}\isamarkuptrue%
\begin{isamarkuptext}%
Note, however, that the user experience of the IDE changes when nesting commands too
deeply. In terms of error handling and failure treatment, there are some noteworthy implementation
differences between the outermost commands and C annotation commands. Naturally, the PIDE toplevel
has been optimized to maximize the error recovery and parallel execution. Inside a command, the
possibilities to mimic this behaviour are somewhat limited. As a workaround useful during development and
debugging, we offer a further pragma for a global annotation, namely \isa{{\isacharasterisk}}
(in complement to the violet \isa{{\isacharat}}), that controls a switch between a
strict and a permissive error handling for nested annotation commands.%
\end{isamarkuptext}\isamarkuptrue%
\begin{isamarkupsubsection*}%
[label = {annot2},type = {scholarly_paper.technical}, args={label = {annot2},type = {scholarly_paper.technical}, Isa_COL.text_element.level = {}, Isa_COL.text_element.referentiable = {False}, Isa_COL.text_element.variants = {{STR ''outline'', STR ''document''}}, scholarly_paper.text_section.main_author = {}, scholarly_paper.text_section.fixme_list = {}, Isa_COL.text_element.level = {}, scholarly_paper.technical.definition_list = {}}]Evaluation Order%
\end{isamarkupsubsection*}\isamarkuptrue%
\begin{isamarkuptext}%
We will now explain why positional languages are affecting the evaluation time of
annotation commands in \csname isaDof.ref\endcsname[type={Isa_COL.figure}]{C-sample3}. This requires a little zoom on how the
parsing is actually executed.

The LALR parsing of our implemented C11 parser can be summarized as a sequence of alternations
between Shift and Reduce actions. By definition of LALR, whereas a unique Shift action is performed
for each C token read from left to right, some unlimited number of Reduce actions are happening
between two Shifts. Internally, the parser manages a stack-like data-structure called
$\alpha^{\text{AST}}$~\isatt{list} representing all already encountered Shift and
Reduce actions (SR). A given $\alpha^{\text{AST}}$~\isatt{list} can be seen as a
\emph{forest of SR nodes}: all leafs are tagged with a Shift, and any other parent
node is a Reduce node. After a
certain point in the parsing history, the top stack element $\alpha^{\text{AST}}$ (cast with the
right type) is returned to \isa{\isacommand{{\isasymsimeq}setup}}.

Since a SR-forest is a list of SR-trees, it is possible to go forward and backward at will in the
actually unparsed SR-history, and execute a sequence of SR parsing steps only when needed. While
every annotation command like \isa{\isacommand{{\isasymsimeq}setup}} is by default attached to
a closest previous Shift leaf, navigation expressions modify the attached node, making the
presentation of $\alpha^{\text{AST}}$ referring to another term focus.

\begin{wrapfigure}{r}{0.62\textwidth}
  \vspace{-22pt}
  \begin{center}
    \includegraphics[width=0.62\textwidth]{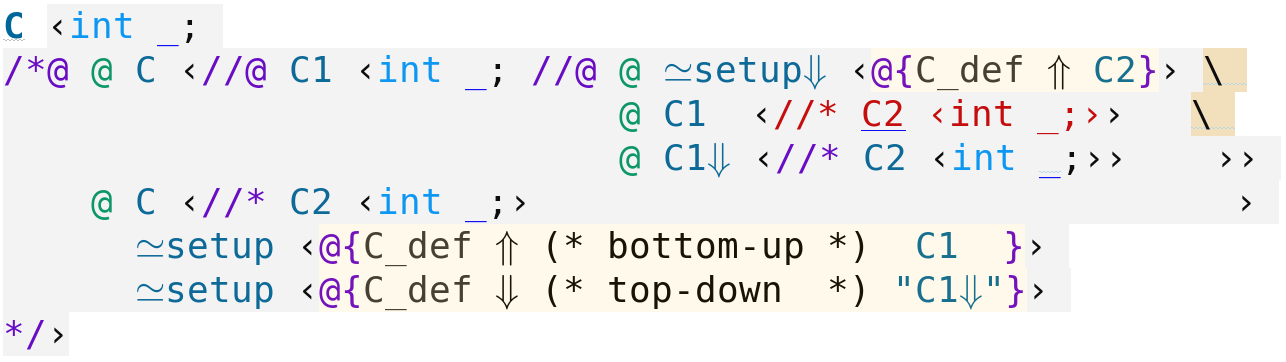}
  \end{center}
  \vspace{-20pt}
\end{wrapfigure}
Instead of visiting the AST in the default bottom-up direction during parsing, it is possible to
store the intermediate results, so that it can be revisited by using another direction strategy, for
example top-down after parsing (where a parent node is executed before any of its children, and
knows how they have been parsed thanks to $\alpha^{\text{AST}}$). This enables commands to decide if
they want to be executed during parsing, or after the full AST has been built. This gives rise to
the implementation of different versions of annotation commands that are executed at different
moments, relative to the parsing process. For example, the
annotation command \isa{\isacommand{{\isasymsimeq}setup}} has been defined for being executed
at bottom-up time, whereas the execution of the variant
\isa{\isacommand{{\isasymsimeq}setup{\isasymDown}}} happens at top-down time. In the above example,
\isa{\isacommand{C{\isadigit{1}}}} is a new command defined by \isatt{C{\char`\_}def}, a
shorthand antiquotation for \isatt{C{\char`\_}Annotation.command{\char`\'}}. Since
\isa{\isacommand{C{\isadigit{1}}}} is meant to be executed during bottom-up time (during parsing), it
is executed before \isa{\isacommand{C{\isadigit{2}}}} is defined (which is directly after parsing).

Note that the C11 grammar has enough scoping structure for the full inference of the C environment
\isatt{env} be at bottom-up time. In terms of efficiency, we use specific
\emph{static} rule wrappers having the potential of overloading default grammar rules (see \csname isaDof.ref\endcsname[type={Isa_COL.figure}]{architecture}), to assign a wrapper to be always executed as soon as a Shift-Reduce
rule node of interest is encountered. The advantage of this construction is that the wrappers are
statically compiled, which results in a very efficient reporting of C type information.%
\end{isamarkuptext}\isamarkuptrue%
\begin{isamarkupsection*}%
[label = {backends},type = {scholarly_paper.technical}, args={label = {backends},type = {scholarly_paper.technical}, Isa_COL.text_element.level = {}, Isa_COL.text_element.referentiable = {False}, Isa_COL.text_element.variants = {{STR ''outline'', STR ''document''}}, scholarly_paper.text_section.main_author = {}, scholarly_paper.text_section.fixme_list = {}, Isa_COL.text_element.level = {}, scholarly_paper.technical.definition_list = {}}]Semantic Back-Ends%
\end{isamarkupsection*}\isamarkuptrue%
\begin{isamarkuptext}%
In this section, we briefly present two integrations of verification back-ends for C. We
chose Clean used
for program-based test generation \cite{DBLP:conf/tap/Keller18}, and AutoCorres \cite{DBLP:conf/pldi/GreenawayLAK14}, arguably the most developed deductive verification environment
for machine-oriented C available at present.

Note that we were focusing on keeping modifications of integrated components minimal, particularly
for the case of AutoCorres. Certain functionalities like position propagation of HOL terms in
annotations, or ``automatic'' incremental declarations~\footnote{\url{https://github.com/seL4/l4v/blob/master/tools/autocorres/tests/examples/Incremental.thy}}
may require internal revisions on the back-end side. This is out of the scope of this paper.%
\end{isamarkuptext}\isamarkuptrue%
\begin{isamarkupsubsection*}%
[label = {clean},type = {scholarly_paper.technical}, args={label = {clean},type = {scholarly_paper.technical}, Isa_COL.text_element.level = {}, Isa_COL.text_element.referentiable = {False}, Isa_COL.text_element.variants = {{STR ''outline'', STR ''document''}}, scholarly_paper.text_section.main_author = {}, scholarly_paper.text_section.fixme_list = {}, Isa_COL.text_element.level = {}, scholarly_paper.technical.definition_list = {}}]A Simple Typed Memory Model: Clean%
\end{isamarkupsubsection*}\isamarkuptrue%
\begin{isamarkuptext}%
Clean (pronounced as: ``C lean'' or ``Céline'' [selin]) is based on a simple,
shallow-style execution model for an imperative target language. It is based on a ``no-frills''
state-exception monad \isa{\isacommand{type{\isacharunderscore}synonym}\ {\isacharparenleft}{\isacharprime}o{\isacharcomma}\ {\isacharprime}{\isasymsigma}{\isacharparenright}\ MON\isactrlsub S\isactrlsub E\ {\isacharequal}\ {\isacartoucheopen}{\isacharprime}{\isasymsigma}\ {\isasymrightharpoonup}\ {\isacharparenleft}{\isacharprime}o\ {\isasymtimes}\ {\isacharprime}{\isasymsigma}{\isacharparenright}{\isacartoucheclose}} with the usual
definitions of \isa{bind} and \isa{unit}. In this language,
sequence operators, conditionals and loops can be integrated. From a concrete program, the
underlying state \isa{{\isacharprime}{\isasymsigma}} is constructed by a sequence of extensible
record definitions:

\begin{enumerate}%
\item Initially, an internal control state is defined to give semantics to
\isa{break} and \isa{return} statements:
\begin{isar}
record control_state =  break_val  :: bool   return_val :: bool
\end{isar}
\isa{control{\isacharunderscore}state} represents the $\sigma_0$ state.

\item Any global variable definition block with definitions $a_1 : \tau_1$ $\dots$ $a_n : \tau_n$  
is translated into a record extension:
\begin{isar}
record \<sigma>$_{n+1}$ = \<sigma>$_n$    +    a$_1$ :: $\tau_1$; ...; $a_n$ :: $\tau_n$
\end{isar}

\item Any local variable definition block (as part of a procedure declaration) 
with definitions $a_1 : \tau_1$ $\dots$ $a_n : \tau_n$ is translated into the record extension:
\begin{isar}
record \<sigma>$_{n+1}$ = \<sigma>$_n$    +    a$_1$ :: $\tau_1$ list; ...; $a_n$ :: $\tau_n$ list; result :: $\tau_{result-type}$ list; 
\end{isar}
where the \isa{list}-lifting is used to model a \emph{stack} of local variable instances
in case of direct recursions and the \isa{result} used for the value of the \isa{return}
statement.%
\end{enumerate}%
\end{isamarkuptext}\isamarkuptrue%
\begin{isamarkuptext}%
The \isa{\isacommand{record}} package creates an
\isa{{\isacharprime}{\isasymsigma}} extensible record type
\isa{{\isacharprime}{\isasymsigma}\ control{\isacharunderscore}state{\isacharunderscore}ext} where the
\isa{{\isacharprime}{\isasymsigma}} stands for extensions that were subsequently ``stuffed'' in
them. Furthermore, it generates definitions for the constructor, accessor and update functions and
automatically derives a number of theorems over them (e.g., ``updates on different fields commute'',
``accessors on a record are surjective'', ``accessors yield the value of the last update''). The
collection of these theorems constitutes the \emph{memory model} of Clean. This
model might be wrong in the sense that it does not reflect the operational behaviour of a particular
compiler, however, it is by construction \emph{logically consistent} since it is
impossible to derive falsity from the entire set of rules.%
\end{isamarkuptext}\isamarkuptrue%
\begin{isamarkuptext}%
\begin{wrapfigure}{r}{0.66\textwidth}
  \vspace{-14pt}
  \includegraphics[width=0.66\textwidth]{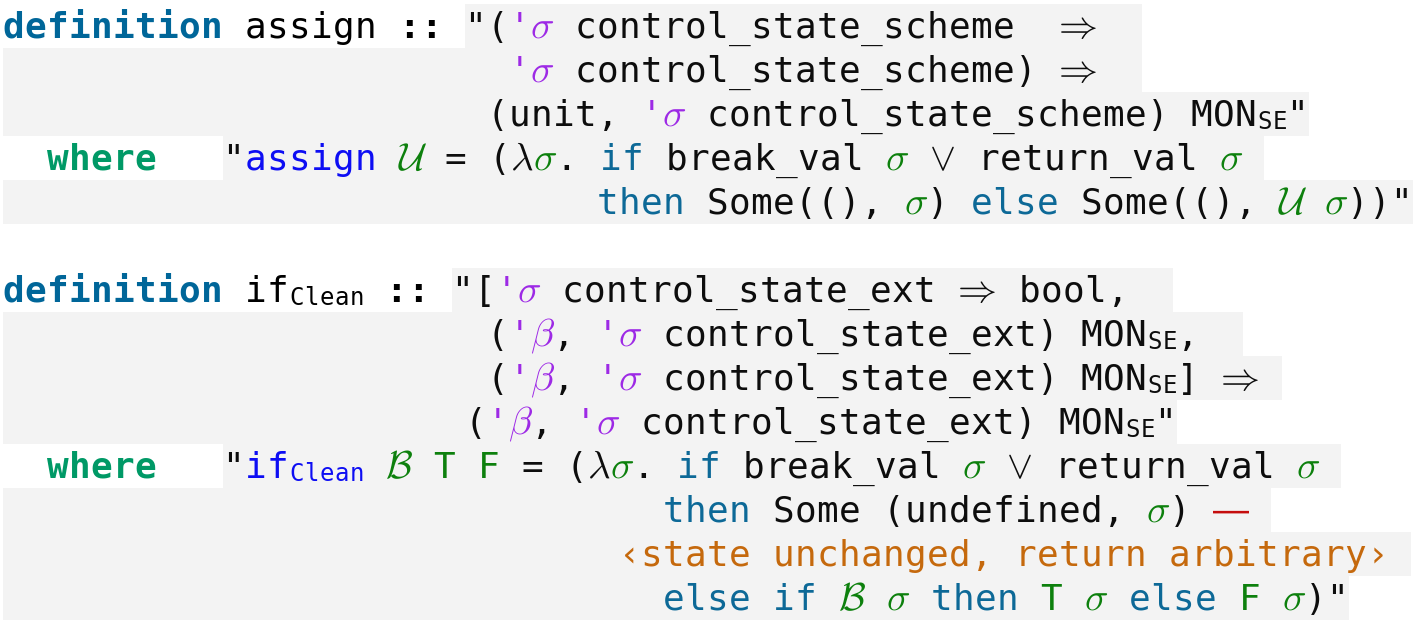}
  \vspace{-30pt}
\end{wrapfigure}
On this basis, assignments, conditionals and loops are reformulated into
\isa{break}-aware and \isa{return}-aware versions as shown in the figure aside. The Clean
theory contains about 600 derived theorems containing symbolic evaluation and Hoare-style
verification rules.

Importing Clean into a theory with its activated back-end proceeds as in \autoref{fig:clean}.
Clean generates for the C program a common type for the state, based on two generated
extensible records --- in the figure: just a global variable \isa{k} and a local variable 
with a stack of result values for \isa{prime\isactrlsub C}. Clean maps machine integers simply 
and naively on the HOL type \isa{int}. The core of this program is represented by two 
generated definitions available subsequently in the logical context, where they are ready to be
used in symbolic executions or proofs.

Generated definitions include push and pop operations for local variable
blocks, for the entire variable space of procedures. Additionally, a specific syntax is introduced to represent assignments on global and local variables. For example,
\isa{i\ {\isacharcolon}{\isacharequal}\ {\isadigit{2}}} internally rewrites to
\isa{assign\ {\isacharparenleft}{\isasymlambda}{\isasymsigma}{\isachardot}\ {\isacharparenleft}{\isacharparenleft}i{\isacharunderscore}upd\ o\ map{\isacharunderscore}hd{\isacharparenright}\ {\isacharparenleft}{\isasymlambda}{\isacharunderscore}{\isachardot}\ {\isadigit{2}}{\isacharparenright}{\isacharparenright}\ {\isasymsigma}{\isacharparenright}}.
The \isa{return} operation is syntactically equivalent to the assignment of the result variable 
in the local state (stack) and sets the \isa{return{\isacharunderscore}val} flag.
On this representation of the C program, the HOL term \isa{prime\isactrlsub C\ n} can be decomposed into  
program test-cases according to a well-established coverage criterion.
Technically, this is done by a variant of the program-based testing method
\begin{isar}
  apply (branch_and_loop_coverage "Suc (Suc (Suc 0))")
\end{isar}
developed in \cite{DBLP:conf/tap/Keller18}, which also uses Clean as semantic basis.
Note that the testing approach does not need the formulation of an invariant,
which is already non-trivial in the given example. 

\begin{figure}
  \centering
  \begin{minipage}{0.53\linewidth}
  \includegraphics[width=0.96\textwidth]{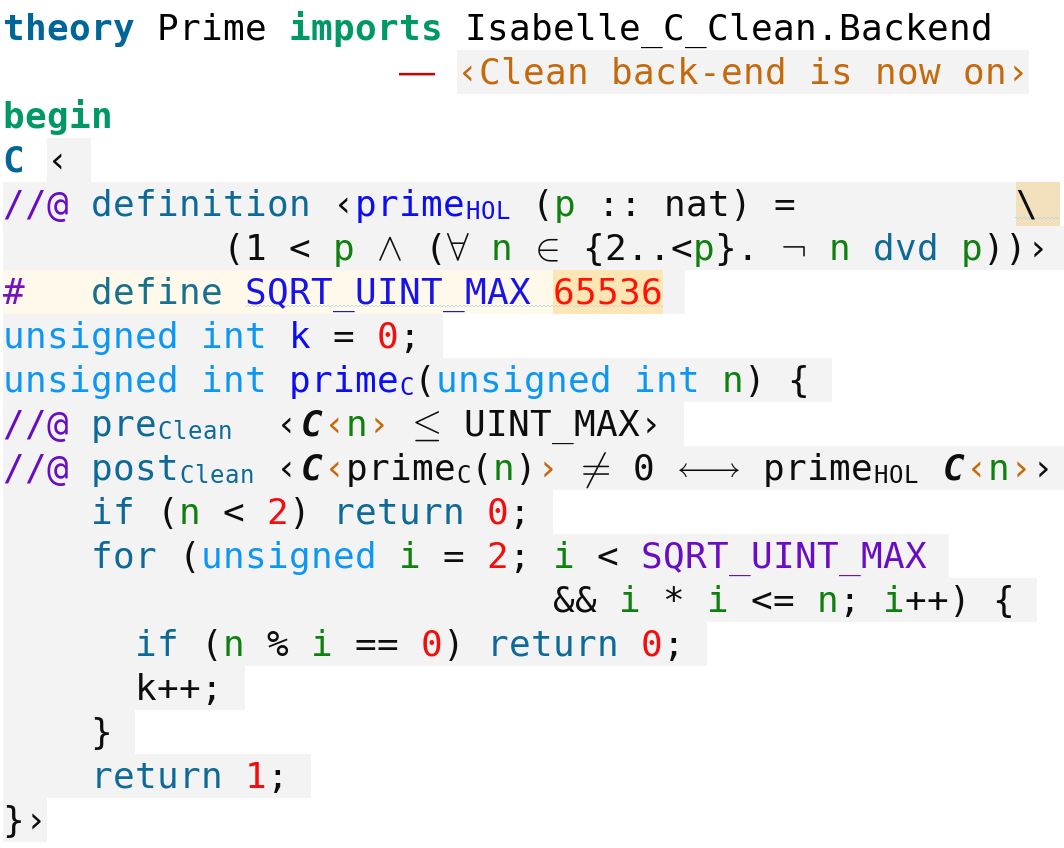}
  \end{minipage}
  \begin{minipage}{0.45\linewidth}
\begin{isar}
prime$_{\text{C}}$_core_def: "prime$_{\text{C}}$_core n \<equiv>
  if$_{\text{Clean}}$ \<Open> (n < 2) \<Close> then return 0 else skip;-
  \<Open> i := 2 \<Close>;-
  while$_{\text{Clean}}$ \<Open> i < SQRT_UINT_MAX \<and> i * i \<le> n\<Close>
    (if$_{\text{Clean}}$ \<Open>n mod i = 0\<Close>
      then return 0 else skip;
     \<Open>k:=k+1\<Close>; assert \<Open> k\<le>UINT_MAX \<Close>
     \<Open>i:=i+1\<Close>; assert \<Open> i\<le>UINT_MAX \<Close>) ;-
  return 1"

prime$_{\text{C}}$_def: "prime$_{\text{C}}$ n \<equiv>
  block$_{\text{Clean}}$ push_local_prime$_{\text{C}}$_state
             (is_prime_core n)
             pop_local_prime$_{\text{C}}$_state"
\end{isar}
  \end{minipage}
  \caption{Activating the Isabelle/C/Clean back-end triggers the generation of theorems}
  \label{fig:clean}
\end{figure}%
\end{isamarkuptext}\isamarkuptrue%
\begin{isamarkuptext}%
\begin{wrapfigure}{r}{0.4\textwidth}
  \vspace{-14pt}
  \includegraphics[width=0.4\textwidth]{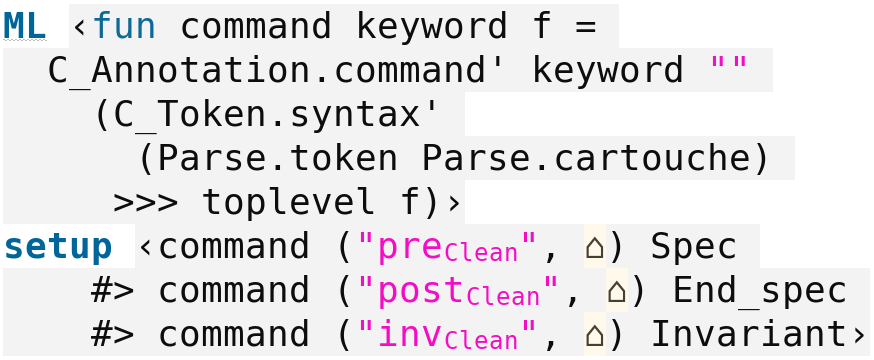}
\end{wrapfigure}
Finally, we will have a glance at the code for the registration of the annotation commands
used in the example. Thanks to
Isabelle/C's function \isatt{C{\char`\_}Annotation.command{\char`\'}}, the registration of
user-defined annotations is very similar to the registration of ordinary commands in the Isabelle
platform.%
\end{isamarkuptext}\isamarkuptrue%
\begin{isamarkupsubsection*}%
[label = {autocorres},type = {scholarly_paper.technical}, args={label = {autocorres},type = {scholarly_paper.technical}, Isa_COL.text_element.level = {}, Isa_COL.text_element.referentiable = {False}, Isa_COL.text_element.variants = {{STR ''outline'', STR ''document''}}, scholarly_paper.text_section.main_author = {}, scholarly_paper.text_section.fixme_list = {}, Isa_COL.text_element.level = {}, scholarly_paper.technical.definition_list = {}}]The Case of AutoCorres%
\end{isamarkupsubsection*}\isamarkuptrue%
\begin{isamarkuptext}%
The AutoCorres environment consists of a C99 parser, compiling to a deepish embedding
of a generic imperative core programming language, over a refined machine word oriented memory
model, and a translator of this presentation into a shallow language based on another Monad for
non-deterministic computations. This translator has been described in \cite{DBLP:conf/pldi/GreenawayLAK14,DBLP:conf/tphol/WinwoodKSACN09} in detail. However, the
original use of AutoCorres implies a number of protocol rules to follow, and is only loosely
integrated into the Isabelle document
model, which complicates the workflow substantially.%
\end{isamarkuptext}\isamarkuptrue%
\begin{isamarkuptext}%
\begin{wrapfigure}{r}{0.48\textwidth}
\vspace{5pt}
\includegraphics[width=0.48\textwidth]{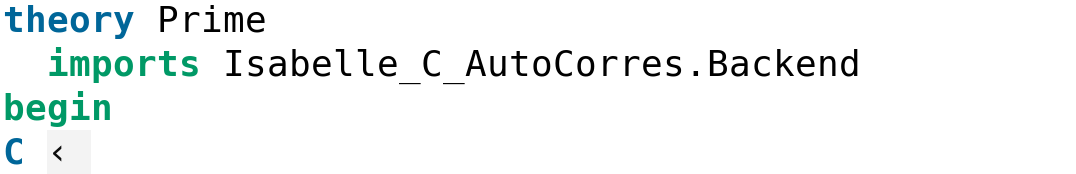}\vspace{-3pt}\hfill\allowbreak%
\vspace{0pt}$\quad\vdots$\vspace{2pt}\hfill\allowbreak%
\includegraphics[width=0.48\textwidth]{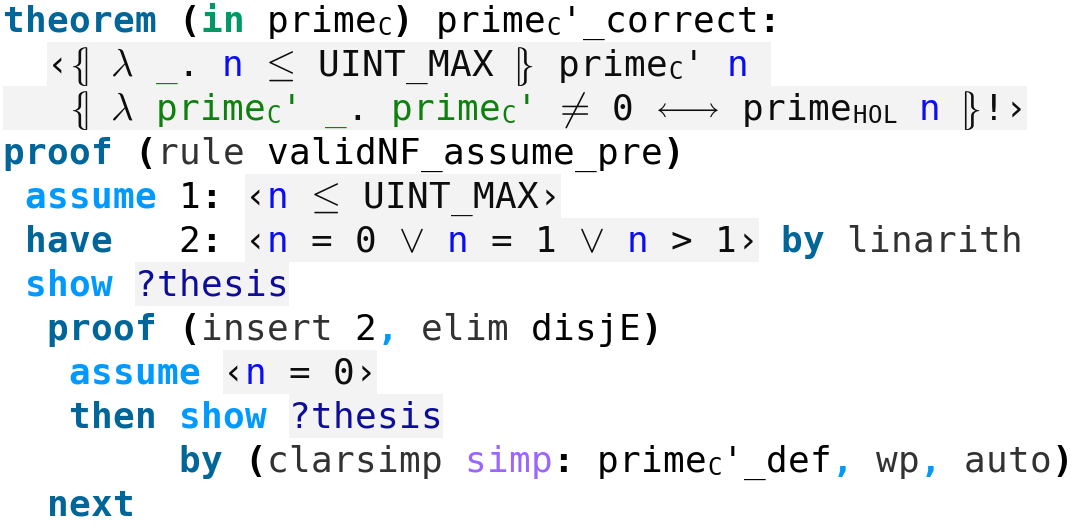}\vspace{-3pt}\hfill\allowbreak%
\vspace{0pt}$\quad\vdots$\vspace{2pt}\hfill\allowbreak%
\end{wrapfigure}
Our running example \isa{prime\isactrlsub C} for Isabelle/C/AutoCorres basically differs in what
the theory is importing in its header. Similarly to Clean, AutoCorres constructs a memory model and
represents the program as a monadic operation on it. Actually, it generates even two presentations,
one on a very precise word-level memory model taking aspects of the underlying processor
architecture into account, and another one more abstract, then it automatically proves the
correspondence in our concrete example. Both representations become the definitions
\isa{prime\isactrlsub C{\isacharunderscore}def} and \isa{prime\isactrlsub C{\isacharprime}{\isacharunderscore}def}. A Hoare-calculus plus
a derived verification generator \isa{wp} from the AutoCorres package
leverage finally the correctness proof.%
\end{isamarkuptext}\isamarkuptrue%
\begin{isamarkuptext}%
Note that the integration of AutoCorres crucially depends on the conversion
$\text{AST}^\text{C11}~\isa{{\isasymRightarrow}}~\text{AST}^\text{C99}$ of $C^\downarrow$
discussed in \csname isaDof.ref\endcsname[type={text}]{parser_validation}. In particular, for the overall seL4
annotations
{\scriptsize\bfseries INVARIANT},
{\scriptsize\bfseries INV},
{\scriptsize\bfseries FNSPEC},
{\scriptsize\bfseries RELSPEC},
{\scriptsize\bfseries MODIFIES},
{\scriptsize\bfseries DONT\_TRANSLATE},
{\scriptsize\bfseries AUXUPD},
{\scriptsize\bfseries GHOSTUPD},
{\scriptsize\bfseries SPEC},
{\scriptsize\bfseries END-SPEC},
{\scriptsize\bfseries CALLS}, and
{\scriptsize\bfseries OWNED\_BY},
we have extended our implementation of $C^\downarrow$ in such a way that the conversion places the
information at the right position in the target AST. Obviously, this works even when navigation is
used, as in \csname isaDof.ref\endcsname[type={Isa_COL.figure}]{C-sample3} left.%
\end{isamarkuptext}\isamarkuptrue%
\begin{isamarkupsection*}%
[label = {concl},type = {scholarly_paper.conclusion}, args={label = {concl},type = {scholarly_paper.conclusion}, Isa_COL.text_element.level = {}, Isa_COL.text_element.referentiable = {False}, Isa_COL.text_element.variants = {{STR ''outline'', STR ''document''}}, scholarly_paper.text_section.main_author = {}, scholarly_paper.text_section.fixme_list = {}, Isa_COL.text_element.level = {}, scholarly_paper.text_section.main_author = {}}]Conclusions%
\end{isamarkupsection*}\isamarkuptrue%
\begin{isamarkuptext}%
We presented Isabelle/C a novel, generic front-end for a deep integration of C11 code into the 
Isabelle/PIDE framework. Based on open-source Lex and Yacc style grammars, we presented a build process 
that constructs key components for this front-end: the lexer, the parser, and a framework for 
user-defined annotations including user-defined annotation commands. 
While the generation process is relatively long, the generated complete library can be loaded in a 
few seconds constructing an environment similar to the usual \isa{\isacommand{ML}} environment for 
Isabelle itself. 20~kLoC large  C sources can be parsed and decorated in PIDE within  seconds.

Our framework allows for the deep integration of the C source into a global document model in which 
literate programming style documentation, modelling as well as static program analysis and verification 
co-exist.  In particular, information from the different tools realized as plugin in the Isabelle 
platform  can flow freely, but based on a clean management of their semantic context and within a 
framework based on conservative theory development. This substantially increases the development agility 
of such type of sources and may be  attractive to  conventional developers, in particular when 
targeting formal certification \cite{DBLP:conf/mkm/BruckerACW18}.

Isabelle/C also forms a basis for future semantically well-understood combinations of back-ends 
based on different semantic interpretations: inside Isabelle, bridge lemmas can be derived 
that describe the precise conditions under which results from one back-end can be re-interpreted
and used in another. Future tactic processes based on these bridge lemmas may open up novel ways 
for semantically safe tool combinations.%
\end{isamarkuptext}\isamarkuptrue%
\paragraph*{Acknowledgments.}
\begin{isamarkuptext}%
The authors warmly thank David Sanán and Yang Liu for encouraging the development and reuse of
$C^\downarrow$, initially started in the Securify project \cite{DBLP:conf/tacas/SananZHZTL17}
({\footnotesize \url{http://securify.sce.ntu.edu.sg/}}).
\end{isamarkuptext}\isamarkuptrue%
\isadelimtheory
\endisadelimtheory
\isatagtheory
\endisatagtheory
{\isafoldtheory}%
\isadelimtheory
\endisadelimtheory
\end{isabellebody}%

\IfFileExists{root.bib}{{\bibliography{root}}}{}

\begin{thebibliography}{10}
\providecommand{\bibitemdeclare}[2]{}
\providecommand{\surnamestart}{}
\providecommand{\surnameend}{}
\providecommand{\urlprefix}{Available at }
\providecommand{\url}[1]{\texttt{#1}}
\providecommand{\href}[2]{\texttt{#2}}
\providecommand{\urlalt}[2]{\href{#1}{#2}}
\providecommand{\doi}[1]{doi:\urlalt{http://dx.doi.org/#1}{#1}}
\providecommand{\bibinfo}[2]{#2}

\bibitemdeclare{inproceedings}{DBLP:conf/itp/AissatVW16}
\bibitem{DBLP:conf/itp/AissatVW16}
\bibinfo{author}{Romain \surnamestart A{\"{\i}}ssat\surnameend},
  \bibinfo{author}{Fr{\'{e}}d{\'{e}}ric \surnamestart Voisin\surnameend} \&
  \bibinfo{author}{Burkhart \surnamestart Wolff\surnameend}
  (\bibinfo{year}{2016}): \emph{\bibinfo{title}{Infeasible Paths Elimination by
  Symbolic Execution Techniques - Proof of Correctness and Preservation of
  Paths}}.
\newblock In: {\sl \bibinfo{booktitle}{Interactive Theorem Proving - 7th
  International Conference, {ITP} 2016, Nancy, France, August 22-25, 2016,
  Proceedings}}, pp. \bibinfo{pages}{36--51},
  \doi{10.1007/978-3-319-43144-4\_3}.

\bibitemdeclare{inproceedings}{DBLP:conf/mkm/BarrasGHRTWW13}
\bibitem{DBLP:conf/mkm/BarrasGHRTWW13}
\bibinfo{author}{Bruno \surnamestart Barras\surnameend},
  \bibinfo{author}{Lourdes Del~Carmen \surnamestart
  Gonz{\'{a}}lez{-}Huesca\surnameend}, \bibinfo{author}{Hugo \surnamestart
  Herbelin\surnameend}, \bibinfo{author}{Yann \surnamestart
  R{\'{e}}gis{-}Gianas\surnameend}, \bibinfo{author}{Enrico \surnamestart
  Tassi\surnameend}, \bibinfo{author}{Makarius \surnamestart Wenzel\surnameend}
  \& \bibinfo{author}{Burkhart \surnamestart Wolff\surnameend}
  (\bibinfo{year}{2013}): \emph{\bibinfo{title}{Pervasive Parallelism in
  Highly-Trustable Interactive Theorem Proving Systems}}.
\newblock In \bibinfo{editor}{Jacques \surnamestart Carette\surnameend},
  \bibinfo{editor}{David \surnamestart Aspinall\surnameend},
  \bibinfo{editor}{Christoph \surnamestart Lange\surnameend},
  \bibinfo{editor}{Petr \surnamestart Sojka\surnameend} \&
  \bibinfo{editor}{Wolfgang \surnamestart Windsteiger\surnameend}, editors:
  {\sl \bibinfo{booktitle}{Intelligent Computer Mathematics - MKM, Calculemus,
  DML, and Systems and Projects 2013, Held as Part of {CICM} 2013, Bath, UK,
  July 8-12, 2013. Proceedings}}, {\sl \bibinfo{series}{Lecture Notes in
  Computer Science}} \bibinfo{volume}{7961}, \bibinfo{publisher}{Springer}, pp.
  \bibinfo{pages}{359--363}, \doi{10.1007/978-3-642-39320-4\_29}.

\bibitemdeclare{proceedings}{DBLP:conf/tphol/2009}
\bibitem{DBLP:conf/tphol/2009}
\bibinfo{editor}{Stefan \surnamestart Berghofer\surnameend},
  \bibinfo{editor}{Tobias \surnamestart Nipkow\surnameend},
  \bibinfo{editor}{Christian \surnamestart Urban\surnameend} \&
  \bibinfo{editor}{Makarius \surnamestart Wenzel\surnameend}, editors
  (\bibinfo{year}{2009}): \emph{\bibinfo{title}{Theorem Proving in Higher Order
  Logics, 22nd International Conference, TPHOLs 2009, Munich, Germany, August
  17-20, 2009. Proceedings}}. {\sl \bibinfo{series}{Lecture Notes in Computer
  Science}} \bibinfo{volume}{5674}, \bibinfo{publisher}{Springer},
  \doi{10.1007/978-3-642-03359-9}.

\bibitemdeclare{misc}{bockenek:hal-02069705}
\bibitem{bockenek:hal-02069705}
\bibinfo{author}{Joshua~A \surnamestart Bockenek\surnameend},
  \bibinfo{author}{Peter \surnamestart Lammich\surnameend},
  \bibinfo{author}{Yakoub \surnamestart Nemouchi\surnameend} \&
  \bibinfo{author}{Burkhart \surnamestart Wolff\surnameend}
  (\bibinfo{year}{2018}): \emph{\bibinfo{title}{{Using Isabelle/UTP for the
  Verification of Sorting Algorithms A Case Study}}}.
\newblock \urlprefix\url{https://easychair.org/publications/preprint/CxRV}.
\newblock \bibinfo{note}{Isabelle Workshop 2018, Colocated with Interactive
  Theorem Proving. As part of FLOC 2018, Oxford, GB.}

\bibitemdeclare{inproceedings}{DBLP:conf/mkm/BruckerACW18}
\bibitem{DBLP:conf/mkm/BruckerACW18}
\bibinfo{author}{Achim~D. \surnamestart Brucker\surnameend},
  \bibinfo{author}{Idir \surnamestart A{\"{\i}}t{-}Sadoune\surnameend},
  \bibinfo{author}{Paolo \surnamestart Crisafulli\surnameend} \&
  \bibinfo{author}{Burkhart \surnamestart Wolff\surnameend}
  (\bibinfo{year}{2018}): \emph{\bibinfo{title}{Using the Isabelle Ontology
  Framework - Linking the Formal with the Informal}}.
\newblock In: {\sl \bibinfo{booktitle}{Intelligent Computer Mathematics - 11th
  International Conference, {CICM} 2018, Hagenberg, Austria, August 13-17,
  2018, Proceedings}}, pp. \bibinfo{pages}{23--38},
  \doi{10.1007/978-3-319-96812-4\_3}.

\bibitemdeclare{article}{DBLP:journals/afp/BruckerTW14}
\bibitem{DBLP:journals/afp/BruckerTW14}
\bibinfo{author}{Achim~D. \surnamestart Brucker\surnameend},
  \bibinfo{author}{Fr{\'{e}}d{\'{e}}ric \surnamestart Tuong\surnameend} \&
  \bibinfo{author}{Burkhart \surnamestart Wolff\surnameend}
  (\bibinfo{year}{2014}): \emph{\bibinfo{title}{Featherweight {OCL:} {A}
  Proposal for a Machine-Checked Formal Semantics for {OCL} 2.5}}.
\newblock {\sl \bibinfo{journal}{Archive of Formal Proofs}}
  \bibinfo{volume}{2014}.
\newblock
  \urlprefix\url{https://www.isa-afp.org/entries/Featherweight\_OCL.shtml}.

\bibitemdeclare{misc}{frama-c-home-page}
\bibitem{frama-c-home-page}
\bibinfo{author}{\surnamestart {CEA}-List\surnameend} (\bibinfo{year}{2019}):
  \emph{\bibinfo{title}{The Frama-C Home Page}}.
\newblock \urlprefix\url{https://frama-c.com}.
\newblock \bibinfo{note}{Accessed \DTMdate{2019-03-24}}.

\bibitemdeclare{inproceedings}{DBLP:conf/tphol/CohenDHLMSST09}
\bibitem{DBLP:conf/tphol/CohenDHLMSST09}
\bibinfo{author}{Ernie \surnamestart Cohen\surnameend}, \bibinfo{author}{Markus
  \surnamestart Dahlweid\surnameend}, \bibinfo{author}{Mark~A. \surnamestart
  Hillebrand\surnameend}, \bibinfo{author}{Dirk \surnamestart
  Leinenbach\surnameend}, \bibinfo{author}{Michal \surnamestart
  Moskal\surnameend}, \bibinfo{author}{Thomas \surnamestart Santen\surnameend},
  \bibinfo{author}{Wolfram \surnamestart Schulte\surnameend} \&
  \bibinfo{author}{Stephan \surnamestart Tobies\surnameend}
  (\bibinfo{year}{2009}): \emph{\bibinfo{title}{{VCC:} {A} Practical System for
  Verifying Concurrent {C}}}.
\newblock In \bibinfo{editor}{Berghofer} et~al.  \cite{DBLP:conf/tphol/2009},
  pp. \bibinfo{pages}{23--42}, \doi{10.1007/978-3-642-03359-9\_2}.

\bibitemdeclare{article}{DBLP:journals/cacm/Earley70}
\bibitem{DBLP:journals/cacm/Earley70}
\bibinfo{author}{Jay \surnamestart Earley\surnameend} (\bibinfo{year}{1970}):
  \emph{\bibinfo{title}{An Efficient Context-Free Parsing Algorithm}}.
\newblock {\sl \bibinfo{journal}{Commun. {ACM}}}
  \bibinfo{volume}{13}(\bibinfo{number}{2}), pp. \bibinfo{pages}{94--102},
  \doi{10.1145/362007.362035}.

\bibitemdeclare{inproceedings}{DBLP:conf/pldi/GreenawayLAK14}
\bibitem{DBLP:conf/pldi/GreenawayLAK14}
\bibinfo{author}{David \surnamestart Greenaway\surnameend},
  \bibinfo{author}{Japheth \surnamestart Lim\surnameend}, \bibinfo{author}{June
  \surnamestart Andronick\surnameend} \& \bibinfo{author}{Gerwin \surnamestart
  Klein\surnameend} (\bibinfo{year}{2014}): \emph{\bibinfo{title}{Don't sweat
  the small stuff: formal verification of {C} code without the pain}}.
\newblock In: {\sl \bibinfo{booktitle}{{ACM} {SIGPLAN} Conference on
  Programming Language Design and Implementation, {PLDI} '14, Edinburgh, United
  Kingdom - June 09 - 11, 2014}}, pp. \bibinfo{pages}{429--439},
  \doi{10.1145/2594291.2594296}.

\bibitemdeclare{article}{DBLP:journals/jfp/Hutton92}
\bibitem{DBLP:journals/jfp/Hutton92}
\bibinfo{author}{Graham \surnamestart Hutton\surnameend}
  (\bibinfo{year}{1992}): \emph{\bibinfo{title}{Higher-Order Functions for
  Parsing}}.
\newblock {\sl \bibinfo{journal}{J. Funct. Program.}}
  \bibinfo{volume}{2}(\bibinfo{number}{3}), pp. \bibinfo{pages}{323--343},
  \doi{10.1017/S0956796800000411}.

\bibitemdeclare{inproceedings}{DBLP:conf/tap/Keller18}
\bibitem{DBLP:conf/tap/Keller18}
\bibinfo{author}{Chantal \surnamestart Keller\surnameend}
  (\bibinfo{year}{2018}): \emph{\bibinfo{title}{Tactic Program-Based Testing
  and Bounded Verification in Isabelle/HOL}}.
\newblock In: {\sl \bibinfo{booktitle}{Tests and Proofs - 12th International
  Conference, {TAP} 2018, Held as Part of {STAF} 2018, Toulouse, France, June
  27-29, 2018, Proceedings}}, pp. \bibinfo{pages}{103--119},
  \doi{10.1007/978-3-319-92994-1\_6}.

\bibitemdeclare{article}{DBLP:journals/tocs/KleinAEMSKH14}
\bibitem{DBLP:journals/tocs/KleinAEMSKH14}
\bibinfo{author}{Gerwin \surnamestart Klein\surnameend}, \bibinfo{author}{June
  \surnamestart Andronick\surnameend}, \bibinfo{author}{Kevin \surnamestart
  Elphinstone\surnameend}, \bibinfo{author}{Toby~C. \surnamestart
  Murray\surnameend}, \bibinfo{author}{Thomas \surnamestart Sewell\surnameend},
  \bibinfo{author}{Rafal \surnamestart Kolanski\surnameend} \&
  \bibinfo{author}{Gernot \surnamestart Heiser\surnameend}
  (\bibinfo{year}{2014}): \emph{\bibinfo{title}{Comprehensive formal
  verification of an {OS} microkernel}}.
\newblock {\sl \bibinfo{journal}{{ACM} Trans. Comput. Syst.}}
  \bibinfo{volume}{32}(\bibinfo{number}{1}), pp. \bibinfo{pages}{2:1--2:70},
  \doi{10.1145/2560537}.

\bibitemdeclare{inproceedings}{DBLP:conf/sosp/KleinEHACDEEKNSTW09}
\bibitem{DBLP:conf/sosp/KleinEHACDEEKNSTW09}
\bibinfo{author}{Gerwin \surnamestart Klein\surnameend}, \bibinfo{author}{Kevin
  \surnamestart Elphinstone\surnameend}, \bibinfo{author}{Gernot \surnamestart
  Heiser\surnameend}, \bibinfo{author}{June \surnamestart
  Andronick\surnameend}, \bibinfo{author}{David \surnamestart Cock\surnameend},
  \bibinfo{author}{Philip \surnamestart Derrin\surnameend},
  \bibinfo{author}{Dhammika \surnamestart Elkaduwe\surnameend},
  \bibinfo{author}{Kai \surnamestart Engelhardt\surnameend},
  \bibinfo{author}{Rafal \surnamestart Kolanski\surnameend},
  \bibinfo{author}{Michael \surnamestart Norrish\surnameend},
  \bibinfo{author}{Thomas \surnamestart Sewell\surnameend},
  \bibinfo{author}{Harvey \surnamestart Tuch\surnameend} \&
  \bibinfo{author}{Simon \surnamestart Winwood\surnameend}
  (\bibinfo{year}{2009}): \emph{\bibinfo{title}{seL4: formal verification of an
  {OS} kernel}}.
\newblock In \bibinfo{editor}{Jeanna~Neefe \surnamestart Matthews\surnameend}
  \& \bibinfo{editor}{Thomas~E. \surnamestart Anderson\surnameend}, editors:
  {\sl \bibinfo{booktitle}{Proceedings of the 22nd {ACM} Symposium on Operating
  Systems Principles 2009, {SOSP} 2009, Big Sky, Montana, USA, October 11-14,
  2009}}, \bibinfo{publisher}{{ACM}}, pp. \bibinfo{pages}{207--220},
  \doi{10.1145/1629575.1629596}.

\bibitemdeclare{article}{DBLP:journals/afp/LammichW19}
\bibitem{DBLP:journals/afp/LammichW19}
\bibinfo{author}{Peter \surnamestart Lammich\surnameend} \&
  \bibinfo{author}{Simon \surnamestart Wimmer\surnameend}
  (\bibinfo{year}{2019}): \emph{\bibinfo{title}{{IMP2} - Simple Program
  Verification in Isabelle/HOL}}.
\newblock {\sl \bibinfo{journal}{Archive of Formal Proofs}}
  \bibinfo{volume}{2019}.
\newblock \urlprefix\url{https://www.isa-afp.org/entries/IMP2.html}.

\bibitemdeclare{inproceedings}{DBLP:conf/fm/LeinenbachS09}
\bibitem{DBLP:conf/fm/LeinenbachS09}
\bibinfo{author}{Dirk \surnamestart Leinenbach\surnameend} \&
  \bibinfo{author}{Thomas \surnamestart Santen\surnameend}
  (\bibinfo{year}{2009}): \emph{\bibinfo{title}{Verifying the Microsoft Hyper-V
  Hypervisor with {VCC}}}.
\newblock In \bibinfo{editor}{Ana \surnamestart Cavalcanti\surnameend} \&
  \bibinfo{editor}{Dennis \surnamestart Dams\surnameend}, editors: {\sl
  \bibinfo{booktitle}{{FM} 2009: Formal Methods, Second World Congress,
  Eindhoven, The Netherlands, November 2-6, 2009. Proceedings}}, {\sl
  \bibinfo{series}{Lecture Notes in Computer Science}} \bibinfo{volume}{5850},
  \bibinfo{publisher}{Springer}, pp. \bibinfo{pages}{806--809},
  \doi{10.1007/978-3-642-05089-3\_51}.

\bibitemdeclare{article}{DBLP:journals/cacm/Leroy09}
\bibitem{DBLP:journals/cacm/Leroy09}
\bibinfo{author}{Xavier \surnamestart Leroy\surnameend} (\bibinfo{year}{2009}):
  \emph{\bibinfo{title}{Formal verification of a realistic compiler}}.
\newblock {\sl \bibinfo{journal}{Commun. {ACM}}}
  \bibinfo{volume}{52}(\bibinfo{number}{7}), pp. \bibinfo{pages}{107--115},
  \doi{10.1145/1538788.1538814}.

\bibitemdeclare{book}{DBLP:books/sp/NipkowPW02}
\bibitem{DBLP:books/sp/NipkowPW02}
\bibinfo{author}{Tobias \surnamestart Nipkow\surnameend},
  \bibinfo{author}{Lawrence~C. \surnamestart Paulson\surnameend} \&
  \bibinfo{author}{Markus \surnamestart Wenzel\surnameend}
  (\bibinfo{year}{2002}): \emph{\bibinfo{title}{Isabelle/HOL - {A} Proof
  Assistant for Higher-Order Logic}}.
\newblock {\sl \bibinfo{series}{Lecture Notes in Computer Science}}
  \bibinfo{volume}{2283}, \bibinfo{publisher}{Springer},
  \doi{10.1007/3-540-45949-9}.

\bibitemdeclare{inproceedings}{DBLP:conf/tacas/SananZHZTL17}
\bibitem{DBLP:conf/tacas/SananZHZTL17}
\bibinfo{author}{David \surnamestart San{\'{a}}n\surnameend},
  \bibinfo{author}{Yongwang \surnamestart Zhao\surnameend},
  \bibinfo{author}{Zhe \surnamestart Hou\surnameend}, \bibinfo{author}{Fuyuan
  \surnamestart Zhang\surnameend}, \bibinfo{author}{Alwen \surnamestart
  Tiu\surnameend} \& \bibinfo{author}{Yang \surnamestart Liu\surnameend}
  (\bibinfo{year}{2017}): \emph{\bibinfo{title}{CSimpl: {A}
  Rely-Guarantee-Based Framework for Verifying Concurrent Programs}}.
\newblock In \bibinfo{editor}{Axel \surnamestart Legay\surnameend} \&
  \bibinfo{editor}{Tiziana \surnamestart Margaria\surnameend}, editors: {\sl
  \bibinfo{booktitle}{Tools and Algorithms for the Construction and Analysis of
  Systems - 23rd International Conference, {TACAS} 2017, Held as Part of the
  European Joint Conferences on Theory and Practice of Software, {ETAPS} 2017,
  Uppsala, Sweden, April 22-29, 2017, Proceedings, Part {I}}}, {\sl
  \bibinfo{series}{Lecture Notes in Computer Science}} \bibinfo{volume}{10205},
  pp. \bibinfo{pages}{481--498}, \doi{10.1007/978-3-662-54577-5\_28}.

\bibitemdeclare{article}{DBLP:journals/afp/TuongW15}
\bibitem{DBLP:journals/afp/TuongW15}
\bibinfo{author}{Fr{\'{e}}d{\'{e}}ric \surnamestart Tuong\surnameend} \&
  \bibinfo{author}{Burkhart \surnamestart Wolff\surnameend}
  (\bibinfo{year}{2015}): \emph{\bibinfo{title}{A Meta-Model for the Isabelle
  {API}}}.
\newblock {\sl \bibinfo{journal}{Archive of Formal Proofs}}
  \bibinfo{volume}{2015}.
\newblock
  \urlprefix\url{https://www.isa-afp.org/entries/Isabelle\_Meta\_Model.shtml}.

\bibitemdeclare{inproceedings}{DBLP:conf/itp/Wenzel14}
\bibitem{DBLP:conf/itp/Wenzel14}
\bibinfo{author}{Makarius \surnamestart Wenzel\surnameend}
  (\bibinfo{year}{2014}): \emph{\bibinfo{title}{Asynchronous User Interaction
  and Tool Integration in Isabelle/PIDE}}.
\newblock In \bibinfo{editor}{Gerwin \surnamestart Klein\surnameend} \&
  \bibinfo{editor}{Ruben \surnamestart Gamboa\surnameend}, editors: {\sl
  \bibinfo{booktitle}{Interactive Theorem Proving - 5th International
  Conference, {ITP} 2014, Held as Part of the Vienna Summer of Logic, {VSL}
  2014, Vienna, Austria, July 14-17, 2014. Proceedings}}, {\sl
  \bibinfo{series}{Lecture Notes in Computer Science}} \bibinfo{volume}{8558},
  \bibinfo{publisher}{Springer}, pp. \bibinfo{pages}{515--530},
  \doi{10.1007/978-3-319-08970-6\_33}.

\bibitemdeclare{inproceedings}{DBLP:journals/corr/Wenzel14}
\bibitem{DBLP:journals/corr/Wenzel14}
\bibinfo{author}{Makarius \surnamestart Wenzel\surnameend}
  (\bibinfo{year}{2014}): \emph{\bibinfo{title}{System description:
  Isabelle/jEdit in 2014}}.
\newblock In \bibinfo{editor}{Christoph \surnamestart
  Benzm{\"{u}}ller\surnameend} \& \bibinfo{editor}{Bruno \surnamestart
  {Woltzenlogel Paleo}\surnameend}, editors: {\sl
  \bibinfo{booktitle}{Proceedings Eleventh Workshop on User Interfaces for
  Theorem Provers, {UITP} 2014, Vienna, Austria, 17th July 2014.}}, {\sl
  \bibinfo{series}{{EPTCS}}} \bibinfo{volume}{167}, pp.
  \bibinfo{pages}{84--94}, \doi{10.4204/EPTCS.167.10}.

\bibitemdeclare{inproceedings}{DBLP:conf/tphol/WinwoodKSACN09}
\bibitem{DBLP:conf/tphol/WinwoodKSACN09}
\bibinfo{author}{Simon \surnamestart Winwood\surnameend},
  \bibinfo{author}{Gerwin \surnamestart Klein\surnameend},
  \bibinfo{author}{Thomas \surnamestart Sewell\surnameend},
  \bibinfo{author}{June \surnamestart Andronick\surnameend},
  \bibinfo{author}{David \surnamestart Cock\surnameend} \&
  \bibinfo{author}{Michael \surnamestart Norrish\surnameend}
  (\bibinfo{year}{2009}): \emph{\bibinfo{title}{Mind the Gap}}.
\newblock In \bibinfo{editor}{Berghofer} et~al.  \cite{DBLP:conf/tphol/2009},
  pp. \bibinfo{pages}{500--515}, \doi{10.1007/978-3-642-03359-9\_34}.

\end{thebibliography}
\end{document}